\documentclass[10pt,journal,compsoc]{IEEEtran}

\usepackage[ruled]{algorithm2e}
\usepackage{amsmath}
\usepackage{amsfonts}
\usepackage{hyperref}
\usepackage{subfig}
\usepackage{multirow}
\usepackage{caption}
\usepackage[shortlabels]{enumitem}
\usepackage[flushleft]{threeparttable} 
\usepackage{supertabular}
\usepackage{adjustbox}
\usepackage{cleveref}
\usepackage{varwidth}

\usepackage{xcolor}

\def\reservedtime{\ensuremath{{t_{V_{r}}}}}
\def\publishedtime{\ensuremath{{t_{V_{p}}}}}
\def\exploitedtime{\ensuremath{{t_{V_{e}}}}}
\def\updatereleasetime{\ensuremath{{t_{U_{r}}}}}
\def\updatedeployedtime{\ensuremath{{t_{U_{d}}}}}

\begin{document}
	
\twocolumn
\label{main:paper}

\title{Software Updates Strategies: a Quantitative Evaluation against Advanced Persistent Threats}

\author{Giorgio Di Tizio,
		Michele Armellini,
        Fabio Massacci
\IEEEcompsocitemizethanks{\IEEEcompsocthanksitem G. Di Tizio (corresponding author) is with University of Trento, Italy.\protect\\
E-mail: giorgio.ditizio@unitn.it
\IEEEcompsocthanksitem M. Armellini is with University of Trento, Italy.
\IEEEcompsocthanksitem F. Massacci is with University of Trento, Italy and Vrije Universiteit Amsterdam, The Netherlands.}% <-this % stops an unwanted space

\thanks{The final version of this paper appears in:  IEEE Transactions on Software Engineering, 2022. DOI 10.1109/TSE.2022.3176674}

}

\IEEEtitleabstractindextext{%
\begin{abstract}
Software updates reduce the opportunity for exploitation. However, since updates can also introduce breaking changes, enterprises face the problem of balancing the need to secure software with updates with the need to support operations. We propose a methodology to quantitatively investigate 
the effectiveness of software updates strategies against attacks of
Advanced Persistent Threats (APTs).
We consider strategies where the vendor updates are the only limiting 
factors to cases in which enterprises delay updates from 1 to 7 months based on SANS data.

Our manually curated dataset of APT attacks
covers 86 APTs and 350 campaigns from 2008 to 2020. 
It includes information about attack vectors, exploited 
vulnerabilities (e.g. 0-days vs public vulnerabilities), and affected software and versions. Contrary to common belief, most APT campaigns employed publicly known vulnerabilities.

If an enterprise could theoretically update 
as soon as an update is released, it would face lower odds of being 
compromised than those waiting one (4.9x) or three (9.1x) months. However, if attacked, 
it could still be compromised from 14\% to 33\% of the times.

As in practice enterprises must do regression testing before applying an update, our major finding is that
one could perform 12\% of all possible updates restricting oneself 
only to versions 
fixing publicly known vulnerabilities without significant changes to
the odds of being compromised compared to a company that updates for all 
versions.
\end{abstract}

}

\maketitle

\IEEEdisplaynontitleabstractindextext

\IEEEpeerreviewmaketitle

\IEEEraisesectionheading{\section{Introduction}\label{sec:introduction}}

\IEEEPARstart{A}{recent} study~\cite{DBLP:conf/ndss/KotziasBVC19} shows that it takes more than 200 days for an enterprise to
align 90\% of their machines with the latest (not known to be 
vulnerable) software version given the need to perform regression testing~\cite{DBLP:conf/ccs/PashchenkoVM20}.

Such behavior is rational because
not all vulnerabilities are always exploited in the wild \cite{DBLP:conf/kdd/BozorgiSSV10}, and several
authors have determined that the actual risk of
slow updates against `mass attackers' is limited
\cite{DBLP:journals/tissec/AllodiM14,DBLP:conf/ccs/BilgeHD17} and often due to specific types of vulnerabilities such as those
traded in Black Markets \cite{DBLP:conf/ccs/Allodi17}
or with other predictable characteristics \cite{jacobsimproving,chen2019using}. Hence, risk analysis might be an effective approach when
considering `mass attackers' which might well be `work averse'
and stick to old exploits until they are no longer profitable~\cite{allodiworkaverse}. However, many companies also face Advanced Persistent Threats (APTs). 
APTs are highly specialized
professionals~\cite{DBLP:conf/fedcsis/RotO17} that use a variety of customized strategies \cite{DBLP:journals/compsec/LemayCMF18}, 
often leveraging on spearphishing \cite{urban13plenty} and 0-days \cite{DBLP:conf/cms/ChenDH14,DBLP:journals/cn/MarchettiPCG16} to maintain a stealthy profile~\cite{DBLP:conf/cms/ChenDH14}. In this scenario, slow updates do not seem appropriate.

Yet, not all the APTs are really \emph{sophisticated}~\cite{DBLP:books/sp/Steffens20}. Some reports challenged some of these `allegations' and observed
that APTs often reuse tools, malware, and
vulnerabilities~\cite{urban13plenty,kaspersky_cve-2015-2545,crowdstrike_cve-2014-0322}.
These reports are based on threat intelligence data with few overlaps~\cite{bouwman2020different} thus capturing only partial information of the APT attacks~\cite{DBLP:conf/uss/LiDPMVS19,DBLP:books/sp/Steffens20}.

These conflicting claims may be due to the lack of a systematic study. Indeed, previous works on APTs
analysis~\cite{DBLP:conf/ciss/UssathJ0M16,DBLP:conf/cms/ChenDH14,DBLP:conf/fedcsis/RotO17,DBLP:conf/IEEEares/VirvilisG13}
mostly reported a \emph{qualitative} analysis of a handful of APTs.
However, relying on qualitative estimations is known to produce risk miscategorization and wrong prioritization~\cite{anthony2008s,kunreuther2002risk} due to several factors like judgmental biases~\cite{kunreuther2002risk}, agenda-setting, and framing~\cite{scheufele2007framing}. Framing of individual reports can produce a distorted perception of the risk. We lack a broad view of the APT landscape that allows companies to correctly assess the advantages and disadvantages of current approaches to software updates.
Data acquisition of APT campaigns, i.e. specific attacks conducted by APT groups, is currently a challenging task. Semi-automated approaches based on report
parsing~\cite{DBLP:conf/esorics/LaurenzaL19} proved to be too riddled
with false positives because the associations between APTs and software vulnerabilities (identified by a CVE) are based on
the presence of keywords and not on the semantics of the document. Here our research questions are as follows:
\begin{enumerate}[label=\emph{RQ\arabic*.},ref=\emph{RQ\arabic*}]
	\item\label{rq:quantitative_analysis} \emph{What are the APTs characteristics that quantitatively describe the landscape of APT campaigns as observable from public reports?}
	\item\label{rq:effectiveness} \emph{Given a quantitative description of both APT campaigns and software updates, how effective are different update strategies to protect against APT campaigns?}
\end{enumerate}

We thus make the following contributions:

	\begin{itemize}
		\item 
		We build a structured, manually verified database in Neo4j of 86 APTs and more
		than 350 campaigns based on an exhaustive search of over 500 technical reports and blogs and up to 22 different resources for each APT. The database~\cite{giorgio_di_tizio_2022_6514817} is available on Zenodo.
		\item \label{rc:methodology} We present a methodology to quantify and compare the effectiveness and cost of software update strategies on historical data about campaigns.
		\item
		We quantitatively evaluated the effectiveness and cost of different software updates strategies, in terms of the \emph{conditional} probability of being compromised and the number of updates required for 5 widely used software products (\emph{Office}, \emph{Acrobat Reader}, \emph{Air}, \emph{JRE}, and \emph{Flash Player}) for the Windows O.S.
	\end{itemize}

\paragraph*{Scope of the work} We provide a quantitative analysis of the risk against APT to allow companies to make rational decisions on software updates. We do not propose new mechanisms to detect and mitigate APTs attacks. 

\section{The Software Update Problem}\label{sec:terminology}
If a company could only update for new functionalities, the choice would be obvious: why fixing what is not broken? Yet, companies must update for security reasons too. However, it is not uncommon that vulnerability fixes are merged with features changes in a single update. Every time a new version of a software is published, one can
\begin{itemize}
	\item \emph{update} immediately;
	\item \emph{wait} some time (e.g. for regression testing) and update;
	\item \emph{skip} the update. 
\end{itemize}
This choice may be influenced by asynchronous events related to the \emph{reservation}, \emph{disclosure}, and \emph{exploitation} of software vulnerabilities in the \emph{current} release.

Unfortunately, a company cannot fully decide \emph{in advance} the configuration they will have when hit (or most frequently not hit) by an attacker as 
it depends on the attacker's choice. A company can only decide on the software updates strategy. To capture what can happen we introduce some terminology.

\subsection{Terminology}
For each vulnerability we identified five instants of time:

\begin{itemize}
	\item \textit{Vulnerability Reserved time (\reservedtime)}: when the CVE entry for the vulnerability is reserved by MITRE; 
	\item \textit{Vulnerability Published time (\publishedtime)}: when the CVE for the vulnerability is published in NVD;
	\item \textit{Vulnerability Exploited time (\exploitedtime)}: when the vulnerability is
	observed to be exploited in the wild;
	\item \textit{Update release time (\updatereleasetime)}: when an update that addresses the vulnerability is
	released.
	\item \textit{Update deployed time (\updatedeployedtime)}: when an update that addresses the vulnerability is deployed.
\end{itemize}

\begin{figure*}
	\captionsetup[subfloat]{labelformat=empty}
	\centering
	\subfloat[At the time when a software update is available, we have 4 cases: \emph{Case 1)} there is no reservation and publication of vulnerabilities before and after the release of an update for the current version. In this case, there is no exploitation of the vulnerability. \emph{Case 2)} after a software update is released, a vulnerability is reserved and disclosed for the current version. \emph{Case 3)} before the release of a software update a vulnerability is reserved for the current version, but the disclosure happens after the update release. \emph{Case 4)} the reservation and disclosure of a vulnerability for the current version happen before the release of an update. Different update strategies can be applied but are all constrained by the presence of a new release. The exploitation events (vertical lines) can happen at any instant of time asynchronously from the reservation-disclosure process and the release of updates. They are classified following Tab.~\ref{tab:vulns_classification}]
	{
		\includegraphics[width=\textwidth,height=0.45\columnwidth]{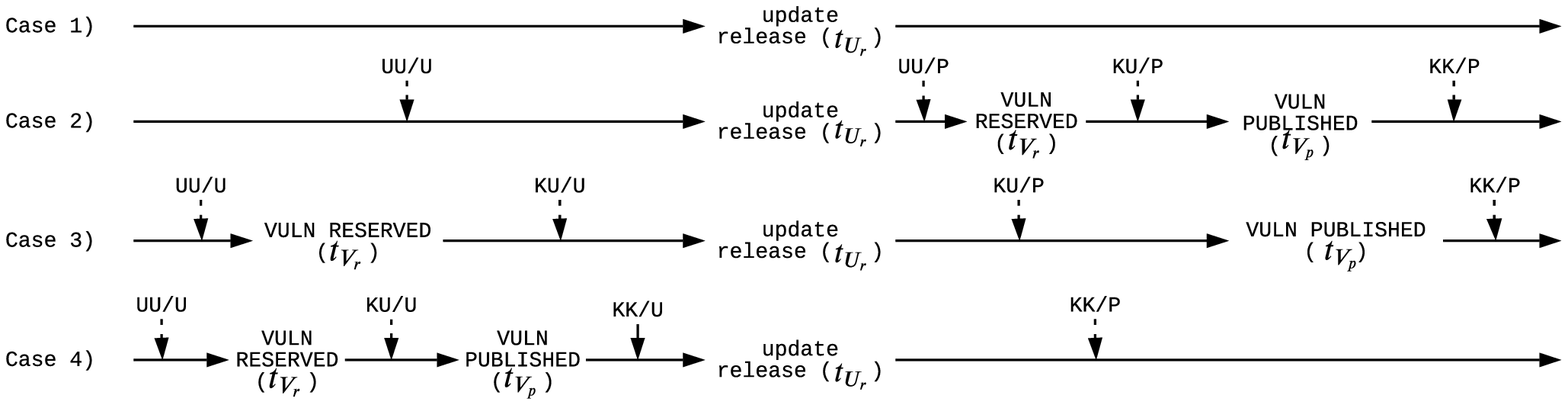}
	}
	\caption{Combinations of vulnerability reservation, disclosure, and exploitation events with the presence of new updates.}
	\label{fig:combination_exploit_update}
\end{figure*}

Tab.~\ref{tab:vulns_classification} shows how we can classify attack scenarios based on the instant of time $\exploitedtime$ and its relative position with the other events: $\reservedtime$, $\publishedtime$, and $\updatereleasetime$
Fig.~\ref{fig:combination_exploit_update} summarizes the possible combinations of the different events. 
\begin{table}
	\captionsetup[subfloat]{labelformat=empty}
	\caption{Classification of Attack Scenarios}
	\label{tab:vulns_classification}
	\subfloat[]{
		\centering
		\begin{tabular}{p{0.25\columnwidth}p{0.65\columnwidth}}
			\hline
			Scenario & Description\\
			\hline\hline
			\emph{Unknown-Unknown/ Unpreventable (UU/U)} & The vulnerability is exploited before a CVE was reserved, before its public disclosure, and before an update for the vulnerability was released.\\\hline
			\emph{Unknown-Unknown/ Preventable (UU/P)} & The vulnerability is exploited before a CVE was reserved, before its public disclosure, but after an update for the vulnerability was released.\\\hline
			\emph{Known-Unknown/ Unpreventable (KU/U)} & The vulnerability is exploited after a CVE was reserved, before its public disclosure, and before an update for the vulnerability was released.\\\hline
			\emph{Known-Unknown/ Preventable (KU/P)} & The vulnerability is exploited after a CVE was reserved, before its public disclosure, and after an update for the vulnerability was released.\\\hline
			\emph{Known-Known/ Unpreventable (KK/U)} & The vulnerability is exploited after a CVE was reserved, after its public disclosure, and before an update for the vulnerability was released.\\\hline
			\emph{Known-Known/ Preventable (KK/P)} & The vulnerability is exploited after a CVE was reserved, after its public disclosure, and after an update for the vulnerability was released.\\
			\hline
		\end{tabular}
	}
\end{table}

\subsection{The Software Update Strategies}
To answer \ref{rq:effectiveness}, we describe the update strategies, summarized in Tab.~\ref{tab:update_strategies}, for an enterprise based on what was discussed previously: \emph{update}, \emph{wait} and then update, or \emph{skip}. It is important to underline that disabling automated updates is not uncommon in enterprise networks~\cite{DBLP:conf/sp/NappaJBCD15,DBLP:conf/uss/XiaoSLLLD18}. This is mainly due to compatibility issues between the updated software and internal projects~\cite{DBLP:conf/ccs/PashchenkoVM20,DBLP:conf/sigsoft/BogartKHT16} that can produce disruption of the enterprise work. In this case delays are introduced to perform regression testing.

We considered the application of a software update with a certain delay, starting from the date on which the strategy bases its decision. We considered different update intervals to determine how the probabilities change if a more responsive approach is employed. We leverage update intervals data from SANS~\cite{sans_patching_2019} based on a variety of enterprises (Government, Financial Services, Healthcare, and Consulting) and from Kotzias et al.~\cite{DBLP:conf/ndss/KotziasBVC19} based on 28k enterprises. Tab.~\ref{tab:patch_interval_sans} shows the update intervals from SANS and maps them to our update strategies.

\begin{table}
	\captionsetup[subfloat]{labelformat=empty}
	\caption{Update strategies}
	\label{tab:update_strategies}
	\subfloat[Each strategy represents an approach for updating the software. The \emph{Immediate} strategy represents the upper bound achievable by an enterprise. The \emph{Planned}, \emph{Reactive}, and \emph{Informed Reactive} are evaluated with different update intervals that represent different level of responsiveness.]{
	\centering
		\begin{tabular}{p{0.15\columnwidth}p{0.15\columnwidth}p{0.55\columnwidth}}
			\hline
			Strategy & Update Interval & Description\\
			\hline\hline
			Immediate & / & Update to each newest version as soon as it is available without any delay\\
			Planned & 1, 3, 7 months & Update to each newest version but wait a delay before the deployment\\
			Reactive & 1, 3, 7 months & Update to the first new (non vulnerable) version only after the publication in NVD of a CVE, wait a delay before the deployment\\ 
			Informed Reactive & 1, 3, 7 months & Update to the first new (non vulnerable) version only after the reservation by MITRE of a CVE entry, wait a delay before the deployment\\ 
			\hline
		\end{tabular}
	}
\end{table}

\par\noindent{\textbf{Immediate strategy:}}
The enterprise updates its software as soon as a new version is available ($\updatereleasetime$) and without delay. If multiple updates are released in the same time interval, the update takes the most recent one. The update is applied even
if a vulnerability for the previous version is not present yet. This is the
theoretical limit for the enterprise because it is bounded only by the release
speed of the vendor. However, this approach is likely impractical because
updates require some time to be deployed in an enterprise to not break
other functionalities.

\begin{table}
	\captionsetup[subfloat]{labelformat=empty}
	\caption{Update Intervals from SANS~\cite{sans_patching_2019}}
	\label{tab:patch_interval_sans}
	\centering
	\subfloat[
	Percentage of enterprises that update weekly, monthly, quarterly, or with other delays. 
	We evaluated the strategies with these update intervals. Enterprises that update weekly are comparable to the \emph{Immediate} strategy. We associated 7 months for the update interval of 'Other' from~\cite{DBLP:conf/ndss/KotziasBVC19}.]
	{
		\begin{tabular}{p{0.12\columnwidth}rp{0.5\columnwidth}}
			\hline
			Update interval & \% Enterprises & Update Strategy Correspondence\\
			\hline\hline
			Weekly & 24.9 & \emph{Immediate}\\
			Monthly & 57.5 & \emph{Planned/Reactive/Informed Reactive} within 1 month delay \\
			Quarterly & 7.7 & \emph{Planned/Reactive/Informed Reactive} within 3 months delay \\ 
			Other & 10.0 & \emph{Planned/Reactive/Informed Reactive} within 7 months delay \\ 
			\hline
		\end{tabular}
	}
\end{table}

\par\noindent\textbf{Planned strategy:}
The company updates its software to each new version with a delay from the release date ($\updatereleasetime$). If multiple updates are released in the same time interval, the company takes the most recent one. This delay factors the time for regression testing and update deployment. The delays are taken from Tab.~\ref{tab:patch_interval_sans}. As in the \emph{Immediate} strategy, the update is \emph{not} triggered by the knowledge of vulnerabilities but only on the availability of a new update.

\par\noindent\textbf{Reactive strategy}
The enterprise updates the software only \emph{after} the publication of a
vulnerability by NVD ($\publishedtime$) with a delay taken from Tab.~\ref{tab:patch_interval_sans}. The new version installed is the first non-vulnerable update available at that time.

\par\noindent\textbf{Informed Reactive strategy}
The enterprise updates the software only \emph{after} the \emph{reservation}
of a vulnerability by MITRE ($\reservedtime$). The new version installed is the first non-vulnerable update available at that time. This strategy describes an enterprise
that pays an annual subscription fee to get information about the non publicly disclosed vulnerabilities from companies that provide
0-days data information (e.g. Exodus Intelligence, Zerodium). The strategy
presents an update interval as the \textit{Reactive} and \textit{Planned} strategies. 

\section{Related Works}\label{sec:relatedworks}
\begin{table}\footnotesize
	\caption{State of the Art \emph{on APTs} - Main Research Topics}
	\label{tab:PaperComparison}
	\centering
	\begin{threeparttable}
	\begin{tabular}{p{0.39\columnwidth}@{}p{0.40\columnwidth}@{}c}
		\hline
		Research Category & State of the Art & This paper\\ 	\hline \hline
		APTs data sources & \cite{DBLP:journals/compsec/LemayCMF18} & \checkmark \\ 

		Metrics for TI sources & \cite{DBLP:conf/uss/LiDPMVS19}, \cite{bouwman2020different} & \\
		Attackers characteristics & \cite{DBLP:conf/cms/ChenDH14}\tnote{*}, \cite{DBLP:conf/ciss/UssathJ0M16}\tnote{*},\cite{urban13plenty} & \checkmark \\
		Detection of attacks & \cite{giura2012context}, \cite{zhao2014extended}, \cite{DBLP:conf/sose/BhattYG14}, \cite{DBLP:conf/icacci/ChandranPP15}, \cite{DBLP:journals/compsec/FriedbergSSF15}, \cite{DBLP:conf/cycon/MarchettiGPC16}, \cite{DBLP:journals/cn/MarchettiPCG16}, \cite{brogi2016terminaptor}, \cite{DBLP:conf/acsac/PeiGSM0ZSZX16}, \cite{DBLP:journals/fgcs/GhafirHPHHRA18}, \cite{DBLP:conf/cdc/SahabanduXC0LP18}, \cite{DBLP:conf/ccs/ShuASSJHR18},\cite{DBLP:conf/sp/MilajerdiGESV19}, \cite{DBLP:conf/uss/ShenS19}, \cite{DBLP:conf/sp/Hassan0M20}, \cite{DBLP:conf/ndss/HanP0MS20},\cite{alsaheelatlas} & \\
		Game Theory & \cite{DBLP:conf/infocom/HuLFCM15}, \cite{riskAPT}, \cite{DBLP:conf/cdc/SahabanduXC0LP18} & \\
		Exploitation likelihood & \cite{DBLP:conf/cycon/MarchettiGPC16}, \cite{riskAPT} & \checkmark \\
		Analysis of update releases &- & \checkmark \\ \hline
	\end{tabular}
	\begin{tablenotes}
	\item[*] Performed high level analysis on few APTs
	campaigns.
\end{tablenotes}
\end{threeparttable}
\end{table}

Tab.~\ref{tab:PaperComparison} shows the research categories addressed by the
state-of-the-art on APTs.
The majority of the research
activity focused on the detection of APTs campaigns while few papers tried to
characterize their behavior, estimate the risk, and evaluate update strategies from real data.

\subsection{APT and Metrics for Threat Intelligence sources}
Lemay et al.~\cite{DBLP:journals/compsec/LemayCMF18} presented a description of
different resources about the activities of more than 40 APTs.
Li et al.~\cite{DBLP:conf/uss/LiDPMVS19} utilized a set of metrics (Volume,
Differential contribution, Exclusive contribution, Latency, Accuracy, and
Coverage) to compare
different public and private Threat Intelligence (TI) data feeds. They observed that in the majority of the data feeds there is no
overlapping of Indicator of Compromise (IOC) and a high number of false positives. Similarly, Bouwman et
al.~\cite{bouwman2020different} analyzed two paid TI and observed very few
overlaps in the indicators for 22 APTs. The distinction is confirmed with a comparison with open TI data. Furthermore, they observed that TI data is employed in the decision process of companies, but there is a lack of metrics to determine the quality of these data.
Several works~\cite{DBLP:conf/ccs/LiaoYWLXB16,DBLP:conf/esorics/LaurenzaL19,DBLP:journals/corr/abs-2104-08618} proposed a (semi-)automated approach based on report parsing to generate a database of IOC. However, merely relying on the results of the automated approach generates many false positives. For
example in~\cite{DBLP:conf/esorics/LaurenzaL19}, we observed that CVEs are wrongly associated to the \textit{admin@338} group in a report about the Poison Ivy malware, where several campaigns from
different actors are described.
We provide a manually curated database from which we can quantitatively evaluate the impact and cost of software update strategies.

\noindent\fbox{\parbox{0.48\textwidth}{\noindent
\emph{Several studies evaluated the overlap among threat data feeds, showing poor accuracy. Mechanisms to semi-automatically extract information from reports are prone to false-positive.}}}

\subsection{Analysis of attackers characteristics}
Ussath et al.~\cite{DBLP:conf/ciss/UssathJ0M16} analyzed 22 reports about APT
campaigns and mapped them into the three phases of an attack (initial compromise, lateral movements, and Command\&Control). They found that most of them employ social engineering
techniques and living-off-the-land techniques. Furthermore, they noted that
0-day vulnerabilities are not exploited frequently by APTs. 
Chen et al.~\cite{DBLP:conf/cms/ChenDH14} studied 4 APT campaigns to analyze the phases
of these attacks and determine possible countermeasures.
Urban et
al.~\cite{urban13plenty} analyzed 93 APT reports (66 different APTs) and determined that
spearphishing is the main attack vector. They then collected OSINT data like domain names and social media information of 30
companies to determine how much information is available to the adversaries.
Additional works on APTs analysis focused on describing the phases of the
attacks and possible countermeasures~\cite{DBLP:conf/fedcsis/RotO17}, the
analysis of the malware employed in a few well-known
campaigns~\cite{DBLP:conf/IEEEares/VirvilisG13}, or the prevalence of living-off-the-land techniques in certain samples~\cite{survivalism}.

To the best of our knowledge, we are the first to analyze more than 350 campaigns exploiting 118 different CVEs from the inspection of more than 500 reports. This massive analysis makes it possible to draw significant conclusions on the efficacy of update strategies.

\noindent\fbox{\parbox{0.48\textwidth}{
\noindent\emph{Although several works provided insights into the APT ecosystem, the analysis focused on a handful of campaigns that make it hard to draw significant conclusions on the characteristics of APTs.}}}

\subsection{Detection of attacks} 
An orthogonal problem is to detect live APTs attacks once they get into the network. Different research proposed to employ machine learning~\cite{DBLP:journals/fgcs/GhafirHPHHRA18,DBLP:conf/icacci/ChandranPP15,alsaheelatlas},
information flow tracking~\cite{brogi2016terminaptor,DBLP:conf/sp/MilajerdiGESV19,DBLP:conf/sp/Hassan0M20,DBLP:conf/ndss/HanP0MS20}, statistical correlation~\cite{DBLP:journals/sadm/SextonSN15}, and big data analysis~\cite{DBLP:conf/cycon/MarchettiGPC16,DBLP:journals/cn/MarchettiPCG16,DBLP:conf/sose/BhattYG14,DBLP:journals/compsec/FriedbergSSF15}.

Shu et al.~\cite{DBLP:conf/ccs/ShuASSJHR18} employed a temporal computational
graph to perform threat hunting activities via graph patterns matching and analyzed
a case study on a DARPA threat detection competition. Pei et
al.~\cite{DBLP:conf/acsac/PeiGSM0ZSZX16} developed a framework to generate a
multi-dimensional weighted graph based on log entries and identify attacks by
the presence of dense connections among logs using unsupervised learning
techniques. They evaluated it over 15 APTs campaigns.

\noindent\fbox{\parbox{0.48\textwidth}{
\noindent\emph{The state-of-the-art focused mainly on the detection and response against APT attacks, while there is a lack of investigation on the orthogonal problem of prevention.}}}
\subsection{Game Theory}Hu et al.~\cite{DBLP:conf/infocom/HuLFCM15} presented a
two-layer attack/defense game to
study APT attackers that make use of insiders and compute the best strategies
for the attacker and defender.
Sahabandu et al.~\cite{DBLP:conf/cdc/SahabanduXC0LP18} formulated a
game-theoretic model to determine the optimal defender strategy in terms of
tracking of information flow (Dynamic Information Flow Tracking). Yang et
al.~\cite{riskAPT} proposed a Nash game to model the response strategy and
minimize the loss of an enterprise against lateral movements in the network of APTs in the
network. We instead focus on the
initial access phase of APTs campaigns and we evaluated the efficacy of software update strategies based on real data of attacks.

\noindent\fbox{\parbox{0.48\textwidth}{
\noindent
\emph{Game theory is extensively applied to find an optimal strategy against targeted attacks. However, these studies employ artificial data and networks.}}}

\subsection{Analysis of exploitation likelihood}
Many works employed ML and statistical methods to analyze vulnerabilities and predict the
exploitation likelihood by joining data from resources like NVD, Exploit DB~\cite{DBLP:conf/kdd/BozorgiSSV10}, historical data on attacks~\cite{DBLP:journals/tissec/AllodiM14,jacobsimproving}, Dark Web forums~\cite{DBLP:conf/cyconus/AlmukayniziNDSS17}, and Twitter~\cite{DBLP:conf/uss/SabottkeSD15,chen2019using}. An extensive discussion of the academic literature on empirical cyber risk can be found in~\cite{woodssystematization}.

Other works investigated actual compromises using logs. Marchetti et al. proposed a framework to prioritize the internal
clients of an organization that are  most likely to be compromised by an APT
using
internal (network logs and flow records) and external (social
media) data~\cite{DBLP:conf/cycon/MarchettiGPC16} and to detect data
exfiltration
using a set of host-based features and flow records
analysis~\cite{DBLP:journals/cn/MarchettiPCG16}.
Similarly, Bilge et al.~\cite{DBLP:conf/ccs/BilgeHD17} and Liu et
al.~\cite{DBLP:conf/uss/LiuSZNKBL15} employed supervised learning algorithms to determine machines at risk of infection from \emph{internal} logs on binary file appearance,
\emph{external} data of misconfigured services (e.g. DNS or BGP), and malicious
behaviors (e.g. spam or phishing).

We extend this line of research by proposing a methodology to evaluate the probability of being compromised by APTs and the cost associated with the update strategy.

\noindent\fbox{\parbox{0.48\textwidth}{
\noindent
\emph{Analysis of historical data about vulnerability and attacks as well as live information provided by logs and social platforms allows one to evaluate the exploitation likelihood.}}}
\subsection{Analysis of update releases}
From the client-side, Nappa et al.~\cite{DBLP:conf/sp/NappaJBCD15} proposed a
systematic analysis of the update process and update delay on client
applications, and performed a survival analysis of vulnerabilities based on data
from Symantec. Similarly, Kotzias et
al.~\cite{DBLP:conf/ndss/KotziasBVC19} presented a longitudinal study of the update behavior for 12
client software and 112 server applications based on data from 28k enterprises.
Sarabi et al.~\cite{DBLP:conf/pam/SarabiZXLD17} employed Symantec dataset to
model users' update delay as a geometric distribution and study 4 different products (Chrome, Firefox, Thunderbird, and
Flash Player).

From the vendor-side, Arora et al.~\cite{DBLP:journals/isr/AroraKTY10} analyzed
vendors' patch behavior as a function of several factors like disclosure time, characteristics of the vendor, and severity of the vulnerability. 
Clark et al.~\cite{DBLP:conf/ccs/ClarkCBS14} studied if agile methods produce
a higher number of vulnerabilities in Firefox. They observed that rapid
software releases do not increase the number of vulnerabilities in the code.
Ozment and Schechter~\cite{DBLP:conf/uss/OzmentS06} analyzed the impact of
legacy code on the number of vulnerabilities observed OpenBSD versions.

Similar to our work, Beres et al.~\cite{beres2008analysing} employed a
discrete-event simulator to determine the exposure reduction produced by
different security policies by varying update speed and mitigations. However, they modeled events like exploits and updates
availability assuming fixed exponential functions looking at global trends observed by a
security firm.

We present a quantitative evaluation of the effectiveness and cost of realistic update strategies by using historical data about APT campaigns.

\noindent\fbox{\parbox{0.48\textwidth}{
\noindent\emph{Several works analyzed the update behavior of clients and vendors. However, there are only theoretical works on the efficacy of updates against targeted attacks for an enterprise.}}}

\section{Methodology}\label{sec:methodology}
In this section, we present a methodology to evaluate the effectiveness and cost of update strategies.

The definition of probabilistic risk assessment~\cite{ezell2010probabilistic} is:
\begin{equation}
\mathit{Risk} = \mathit{Pr(Compr|Attack)} \cdot \mathit{P(Attack)} \cdot \mathit{Impact}\label{eq:risk_def}
\end{equation}
How to determine $\mathit{P(Attack)}$ is still an unsolved problem in cyber-security~\cite{allodi2017security} while the $\mathit{Impact}$ of cyber-attacks has received extensive discussion~\cite{anderson2019measuring,DBLP:conf/sp/DambraBB20}.
In this paper, we focus on $\mathit{P(Compr|Attack)}$ i.e. the \emph{conditional} probability of being compromised given an attack (or campaign as used in this paper). We propose a methodology to compute the \emph{conditional} probability of being compromised in Eq.~\ref{eq:risk_def} by employing historical data about releases available, vulnerabilities, and their exploitation in campaigns.
Tab.~\ref{table:methodology} overviews our methodology.

\begin{table*}
	\caption{Methodology overview}
	\label{table:methodology}
	{\begin{tabular}{rp{450pt}}
			\hline
			\hline
			\multicolumn{2}{l}{\textbf{Step 1: Extract APT and software data}} \\
			INPUT & APT groups from MITRE Att\&ck\\
			OUTPUT & A set of campaigns in the form $<$\textit{APT\_name,CVE,date}$>$, $<$\textit{APT\_name,attack\_vector,date}$>$ and a set of software updates in the form $<$\textit{sw,update,release\_date}$>$\\
			PROCEDURE & Identify campaigns information and software releases:
			\begin{itemize}
				\item Collect resources describing campaigns for each APT based on Threat Actor Encyclopedia~\cite{THAICERT-ENCICLOPEDIA} and Internet searches using the MITRE APT name and "CVE" as keywords;
				\item Manually extract from resources the key information: \emph{date} when campaign is observed, \emph{CVE(s)} exploited, \emph{attack\_vector} employed;
				\item For each CVE, automatically extract software and versions affected from NVD;
				\item Manually extract from software vendors website the \emph{update} number and  \emph{date} of release.
			\end{itemize}
			\\
			\hline
			\multicolumn{2}{l}{\textbf{Step 2: Instantiate update strategy}} \\
			INPUT & A set of software updates ($<$\textit{sw,update,release\_date}$>$), update strategy (\emph{Immediate}, \emph{Planned}, \emph{Reactive}, \emph{Informed Reactive}), CVEs exploited in APT campaigns\\
			OUTPUT & A matrix that describes the application of updates for the software in the period [2008-2020]\\
			PROCEDURE & Create a matrix with rows identifying software versions and columns identifying months in [2008-2020] that determines the installed software version at a given time:
			\begin{itemize}
				\item Select the entry corresponding to the first vulnerable version available on 01/2008 (same for all strategies);
				\item Select another entry corresponding to a new version depending on the update strategy: on the release date of an update for the software (\emph{Immediate}) or with a delay (\emph{Planned}), on the publication (\emph{Reactive}) or reservation date (\emph{Informed Reactive}) of a CVE for the software with a delay;
				\item Consider availability of non-vulnerable updates at the time of publication of a CVE when computing delay for \emph{Reactive} and \emph{Informed Reactive}.
			\end{itemize}
			\\
			\hline
			\multicolumn{2}{l}{\textbf{Step 3: Instantiate APT campaigns events}} \\
			INPUT & Set of events for different campaigns ($<$\textit{APT\_name,CVE,date}$>$)\\
			OUTPUT & A set of matrices of campaigns. Each matrix describes the software versions targeted by a certain campaign in the period [2008-2020]\\
			PROCEDURE &  For each campaign, create a matrix with rows identifying software versions and columns identifying months in the [2008-2020] that determines targeted software version at a given time:
			\begin{itemize}
				\item Extract the affected software versions from the CVEs;
				\item Select the entry of the affected software versions from the \emph{date} of the campaign up to 2020.
			\end{itemize}
			\\
			\hline
			\multicolumn{2}{l}{\textbf{Step 4: Generate pessimistic scenarios}} \\
			INPUT & A matrix that describes the application of updates for the software in the period [2008-2020]\\
			OUTPUT & A matrix that describes the application of updates for the software in the period [2008-2020] and maintains both versions during the transition month\\
			PROCEDURE & Update matrix to maintain the previous version in the month in which a new update is installed:
			\begin{itemize}
				\item For each month in which the software version is updated to a new version, keep the entry corresponding to the previous version installed for that month only.
			\end{itemize}
			\\
			\hline
			\multicolumn{2}{l}{\textbf{Step 5: Compute conditional probability of being compromised}} \\
			INPUT & A set of matrices of update strategies and a set of matrices of campaigns events\\
			OUTPUT & The conditional probability of being compromised given a set of campaigns are targeting you ($\mathit{P(Compr|Attack)}$) based on the update strategy, \# of updates performed\\
			PROCEDURE & Compute successful campaigns targeting installed software in the period [2008-2020]. For each matrix of update strategy:
			\begin{itemize}
				\item Select a matrix of campaign events and compute the element-wise product of the matrix with the update strategy matrix to identify the intersection of installed and targeted software versions;
				\item Sum rows of the resulting matrix to determine the months when a campaign is successful, save the campaign if successful;
				\item Continue with another matrix of campaign events until no more campaigns.
				\item Compute conditional probability as the number of successful campaigns divided by the number of matrix campaigns considered;
				\item Compute the number of updates counting non-empty rows in the matrix of update strategy.
			\end{itemize}
			\\
			\hline
			\multicolumn{2}{l}{\textbf{Step 6: Compare strategies effectiveness}} \\
			INPUT & The successful campaigns for the different update strategies\\
			OUTPUT & Confidence Intervals (CI) for update strategies\\
			PROCEDURE & Compare the CI intervals of different update strategies:
			\begin{itemize}
				\item Compute the Agresti-Coull 95\% CI for the proportion of successful campaigns by update strategy;
				\item Compare intervals, if they overlap update strategies are similar;
				\item Compute pair-wise agreement of successful campaigns for pair of update strategies and Agresti-Coull 95\% CI for the resulting proportion of agreement. The interval identifies the expected range of proportion of campaigns that succeed against both update strategies.
			\end{itemize}
			\\
			\hline
			\hline
		\end{tabular}
	}
\end{table*}

\subsection*{Step 1: Extract APT and software data}
To collect data we analyzed both unstructured (technical reports, blogs about APT campaigns, and vendor's repositories) and structured (MITRE Att\&ck and NVD repositories) public sources.

\par\noindent\textbf{Unstructured sources:}
Similar to Urban et al.~\cite{urban13plenty}, we manually collected data
about APT campaigns from more than 500 technical reports and blogs.
We started from the MITRE Att\&ck APT groups list and one researcher:
\begin{itemize}
\item collected the reports associated with each APT group from the Threat Actor Encyclopedia~\cite{THAICERT-ENCICLOPEDIA}, that relies on sources like Malpedia, MISP, AlienVault, and MITRE;
\item extended this set of resources by searching on the Internet for reports using as keywords the APT name as stated in MITRE (e.g. Stealth Falcon) and the term "CVE" until data saturation was reached, i.e. new reports do not add new information to the APT campaigns. The reports are obtained from cyber-security companies like Kaspersky, FireEye, Palo Alto Networks, Google Project Zero as well as from technical forums and blogs.
\end{itemize}

\subsubsection*{Extracted Information}
Two researchers independently analyzed the
content of each report manually to identify the following information for a campaign: 
\begin{itemize}
	\item the \emph{date} when the campaign is first observed;
	\item the \emph{CVE(s)} exploited;
	\item the \emph{attack vector(s)} employed.
\end{itemize}
We uniquely identify a campaign using the \emph{date} in which it is first observed. If a campaign employs different attack vectors and/or different CVEs, we create multiple entries in the form $<$\textit{APT\_name,attack\_vector,date}$>$ or $<$\textit{APT\_name,CVE,date}$>$. Each entry is linked to one or more reports containing this information.

We do not perform open coding because the information in the reports is deterministic and already based on the MITRE industry standards on CVEs~\footnote{\url{https://www.cve.org/About/History}} and Initial Access Tactic\footnote{\url{https://attack.mitre.org/tactics/TA0001/}}.
The association of CVEs and attack vectors to a certain APT is based on the explicit attribution in the consulted resources.
Let us consider the following snippet from a Mandiant report referring to APT12\footnote{\url{https://www.mandiant.com/resources/darwins-favorite-apt-group-2}}:
	\begin{center}
	\noindent\fbox{%
		\parbox{0.48\textwidth}{%
			{\fontfamily{courier}\selectfont
				In June 2014, the [Arbor Networks] blog highlighted that the backdoor was utilized in campaigns from March 2011 till May 2014.
				Following the release of the article, FireEye observed a distinct change in RIPTIDE’s protocols and strings. \ldots FireEye dubbed this new malware family HIGHTIDE.\\
				On Sunday August 24, 2014 we observed a spear phish email sent to a Taiwanese government ministry. Attached to this email was a malicious Microsoft Word document (MD5: f6fafb7c30b1114befc93f39d0698560) that exploited CVE-2012-0158. It is worth noting$\ldots$}
			}%
	}
\end{center}
we extracted the following information: \emph{date}=08/2014; \emph{CVE}=CVE-2012-0158; \emph{attack vector}=spearphishing attachment.

The entries were then reviewed by a third researcher, not involved in 
the initial manual analysis, to resolve inconsistencies. Cohen's kappa 
values are 1, 0.976, and 0.863 for the CVE, date, and attack vector 
respectively (42 disagreements over 652 entries) which show a good 
agreement among the raters. Total agreement on CVEs is 
unsurprising as CVEs are unique strings and reported by copying and 
pasting the string into the data collection form. Such agreement would 
not happen between a manual rater and an automatic procedure as 
we already noted for DAPTSET \cite{DBLP:conf/esorics/LaurenzaL19} 
which is so riddled with false positives to be unusable. Simply, an 
automatic procedure will collect all CVEs including those that a 
human rater will see as clearly irrelevant (past campaign, related examples, etc.).
 Most disagreements are on the attack vector as the mapping of 
the natural description into the corresponding MITRE Att\&ck category 
is sometimes amenable to interpretation (27 out of 
the 42 disagreements).

To resolve uncertainty among resources, we made the following \emph{conservative} assumptions:
\begin{itemize}
	\item if report A says CVE-1 is exploited by an APT campaign and report B says CVE-2 is exploited we mark both CVE-1 and CVE-2 as exploited by the APT in question. 
	\item if report A says an APT campaign started on month X and report B says an APT campaign started on month Y we mark them as two distinct campaigns.
\end{itemize}
It is not uncommon that different security companies have non-overlapping information about APT campaigns~\cite{DBLP:conf/uss/LiDPMVS19,bouwman2020different}. We discuss the implications of this choice in \S\ref{sec:limitations}. 
Fig.~\ref{fig:reports_APT} in \S\ref{sec:dataset} summarizes the number of reports per APT.

For the software, we retrieved versions for a subset of the targeted 
products (discussed in \S\ref{sec:sec:client-app}) with their release date. This 
procedure was manual as it 
is not trivial to obtain past versions release date~\cite{DBLP:conf/sp/NappaJBCD15} because vendors' repositories are unstructured 
and not intended for past versions indexing.

\par\noindent\textbf{Structured sources:}
For each CVE obtained from the unstructured sources, we automatically extracted the list
of products and versions affected from NVD. This information
is integrated with the Common
Platform Enumeration (CPE) Dictionary. From the CPE, we extracted
the list of vulnerable versions based on the CPE Match Strings.\footnote{For example, CVE-2016-4113 affects all the versions of \textit{Flash Player} up
to 21.0.0.213. The associated JSON NVD file does not provide the entire list of affected
versions (including the updates) in the CVE description but a CPE URI of the
form \textit{cpe:2.3:a:adobe:flash\_player:*:*:*:*:*:*:*:*",
	"versionEndIncluding":"21.0.0.213"}, we thus matched the CPE in the CPE
dictionary to get the list of all prior versions affected.}

\subsection*{Step 2: Instantiate update strategies}
We employed a matrix representation to compare update strategies and APT campaigns.

Each strategy is represented as a matrix in which the rows represent the versions
of the different products (e.g. \emph{Acrobat Reader} 9.2, \emph{Flash Player 11.0.1.152}) and the column a specific date with a month-base granularity (e.g. 12/2009). A matrix cell is 1 if, on that date, that version of the product is installed and otherwise 0. For a first approximation, we avoid considering the presence of multiple versions installed for the same product\footnote{We assume an update is applied on enterprise's machines
	at once.}.

All strategies start from the same version, that is the oldest vulnerable version of a campaign that
is available at the beginning of 2008. A strategy updates its version based on the
release date of a new version, the publication date, and the reservation date of
a CVE for the \textit{Immediate} and \textit{Planned},
\textit{Reactive}, and \textit{Informed Reactive} strategy respectively.

The first two strategies (\textit{Planned} and \textit{Immediate}) update 
when a new release for the software is available (w/ and w/o an update 
interval respectively). If multiple software versions are released on the 
same date, they will update to the newest consistent version.\footnote{For example, if the current \emph{JRE} version installed is 
\emph{6u6} and a new update for JRE \emph{5u13} is released after 
that, the update is ignored because it represents a downgrade of a 
major update.}
	
	For the latter two
	strategies (\textit{Reactive} and \textit{Informed Reactive}) the next version installed, if available, is the \emph{first most
		recent version} that is not affected by the CVE. We also considered the availability of updates based on the attack scenario in \emph{Step 4}.

\subsubsection*{Identification of the outcome of attack scenarios}\label{sec:sec:attack_scenario}
Depending on the availability of an update at
the time of the publication of the CVE we have to discern two scenarios:
\begin{itemize}
	\item The release of an update is available \emph{before} the publication of the CVE ($\updatereleasetime \le \publishedtime$). The time when a company may decide to update because it is aware of the vulnerability is correctly computed from the time when a new vulnerability is published. This is the
	(implicit) assumption in~\cite{DBLP:conf/ndss/KotziasBVC19,DBLP:conf/sp/NappaJBCD15}.
	\item The release of an update is available \emph{after} the publication of the CVE ($\publishedtime<\updatereleasetime$).
	In this case, computing the time when a company may decide to update from the time of publication of the vulnerability will
	include an interval of time where a vulnerability for the version of a
	product is known but a non-vulnerable version has not been released yet
	($\updatereleasetime - \publishedtime$).
	The time available to the company must be computed from the time when the release is available.
\end{itemize}
\subsection*{Step 3: Instantiate APT campaigns events}
We created a matrix for each APT campaign with the same rows and columns of
the update strategy matrix in \emph{Step 2}. An entry is set to 1 if the version is affected by a CVE exploited by the campaign from the date when the campaign starts until 2020.

\subsection*{Step 4: Generate pessimistic scenarios}
The updates and attacks have a month-based granularity because most of the resources \emph{do not} contain information about the exact day in the month in which an update is published or a campaign is performed. We further discuss the limitation of these data in \S\ref{sec:limitations}.

To balance possible interleaves between updates and campaigns within the same month, we performed two analyses: a pessimistic \emph{APT-first} scenario and an optimistic \emph{Update-first} scenario, that assume the campaign is executed before or after the update respectively. 

To simulate the \emph{APT-first} scenario, we create a new matrix from the update strategy matrix where we maintained the previous version also in the month in which the new update is installed. In other words, the two versions coexist in the month. Thus, we simulate the application of the update later in the month while allowing the APT to exploit the vulnerability. This is done by keeping selected the entry corresponding to the previous version also in the column in which we move to another version for each update strategy matrix generated in the \emph{Step 2}.

\subsection*{Step 5: Compute conditional probability of being compromised}

We evaluate at each instant of time, with a \emph{month-base} granularity, the sequence of versions installed on a set of software products for each strategy and compare them with the software exploited by the APTs to determine the potentially successful campaigns.\footnote{For example, in 12/2009 the CVE-2009-4324, affecting \textit{Acrobat Reader} up to version 9.2, is exploited in the wild. If at any time from 12/2009 an update strategy updates to one of these versions, then the campaign is potentially successful.} We use the term \emph{potentially successful} because the success of exploiting a vulnerability depends on the characteristics of the execution environment~\cite{DBLP:conf/uss/DashevskyiSMS14}. A campaign is considered successful if it exploits at least one of the software products considered.
From the matrix of updates obtained from \emph{Step 2,5} and the matrix of campaigns obtained from \emph{Step 3}, we compute the \emph{conditional} probability of being compromised given one is targeted by the campaigns at a given instant of time $t_{i}$. The probability is computed as the number of potentially successful campaigns at time $t_{i}$ over the total number of campaigns active at that instant of time.
\begin{equation}
P(Compr|C,t=t_{i}) =
\frac{\mathit{|potentially~successful~campaigns_{t_{i}}|}}{\mathit{|active~campaigns_{t_{i}}|}}\label{eq:prob_compromise}
\end{equation}
where:
\begin{itemize}
	\item $\mathit{|potentially~successful~campaigns_{t_{i}}|}$ is the number of active
	campaigns at time $t_{i}$ that exploit at least one version of a product currently installed at that time.
	\item $\mathit{|active~campaigns_{t_{i}}|}$ is the total number of active campaigns at time $t_{i}$.
\end{itemize}
We computed the $\mathit{|potentially~successful~campaigns_{t_{i}}|}$ by performing an element-wise product of the matrix of update strategy with each matrix describing an APT campaign. The resulting matrix identifies the versions that were installed \emph{and} exploited by the campaign in a given month. With the sum of the rows of the resulting matrix, one obtains a vector of values $\ge$ 0 for each $\mathit{t_{i}}$.\footnote{Values can be $>$ 1 if campaigns can exploit different products.} If an entry at time $\mathit{t_{i}}$ is $>$ 0, then the campaign is included in $\mathit{|potentially~successful~campaigns_{t_{i}}|}$.

The overall percentage of potentially successful campaigns over the total number of campaigns in the entire interval of time is computed as:
\begin{equation}\label{eq:overall_prob_compromise}
\resizebox{\columnwidth}{!}{$
\begin{split}
P(Compr|C) & =
\frac{\mathit{|potentially~successful~campaigns|}}{\mathit{|campaigns|}}\\
& = \frac{\mathit{|\{C | \exists t_{i} : C \in potentially~successful~campaigns_{t_{i}} \}|}}{\mathit{|campaigns|}} 
\end{split}
$}
\end{equation}
In other words, the total number of potentially successful campaigns is obtained from the set of campaigns that could be successful in at least one instant of time $t_i$. If a campaign can succeed in several instants of time, it is counted only once in the period of interest.

The number of software updates is obtained from the matrix representing the strategy, by counting the number of rows that contain at least one non-zero entry in the columns.

\subsection*{Step 6: Compare strategies effectiveness}
For each update strategy, we obtain from Eq.~\ref{eq:overall_prob_compromise} a probability of being compromised based on the sample of campaigns considered. To predict the range in which the probability of being compromised for the entire population of campaigns resides we compute a confidence interval (CI). In case of binary outcomes (success, failure), we compute the Agresti-Coull confidence interval~\cite{agresti1998approximate} that is recommended when the sample size is $\ge$ 40~\cite{10.1214/ss/1009213286}.
From the CIs of different update strategies, we can then compare their performance. Two strategies are similar if their CIs significantly overlap.

We then determine the percentage of campaigns for which the two strategies behave in the same way by computing the proportion of campaigns that either succeeded or failed against both strategies. By computing the Agresti-Coull interval for the resulting proportion we obtain the range of similarity of the two strategies in terms of the percentage of campaigns that both succeed or failed against two update strategies.

\section{Dataset}\label{sec:dataset}

We considered only APT groups that launched at least one campaign from 2008 to 01/2020 and for which a precise date for the campaign is present in at least one report.

\begin{figure}
	\captionsetup[subfloat]{labelformat=empty}
	\centering
	\subfloat[Only for 11 APTs (~13\%) we were not able to find more than one resource for their campaigns. This is typically due APTs that are not particularly active or that are tracked by a single cyber-security company.]
	{
		\includegraphics[width=\columnwidth,height=0.6\columnwidth]{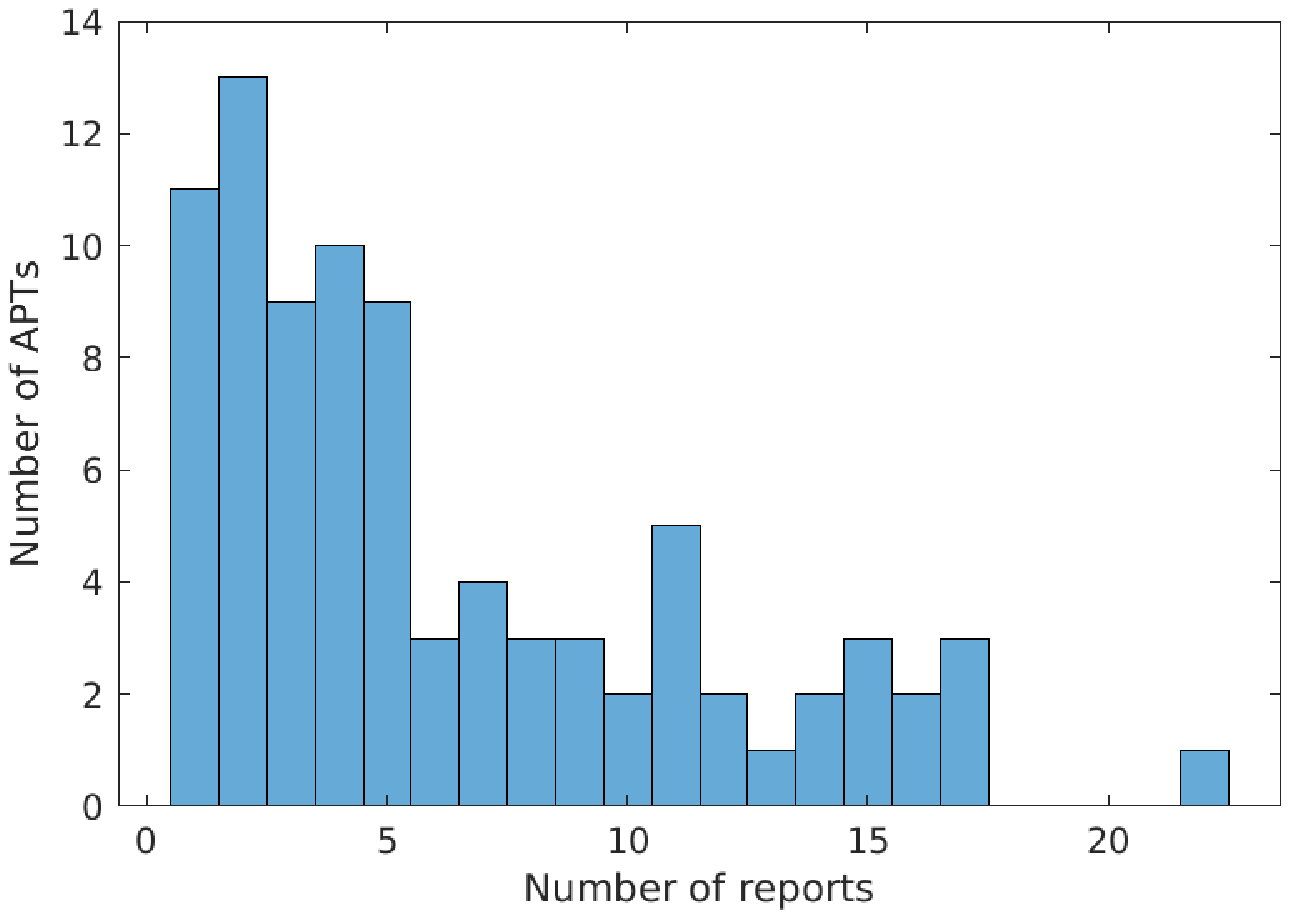}
	}
	\caption{Number of collected reports per APT.}
	\label{fig:reports_APT}
\end{figure}

The final database contains information about 86 APT groups. For the excluded APTs,
we either did not find information for their campaigns, or the date of their
campaign was not known. For example, the Kaspersky article~\cite{kasperskyEquation} provides a list of CVEs but does not provide information about the campaign when they were exploited. Fig.~\ref{fig:reports_APT} shows the distribution of reports per APT.

\noindent\fbox{\parbox{0.48\textwidth}{\noindent\emph{For more than half of the APT campaigns saturation is reached with at most 5 distinct resources, while for some APTs we collected more than 15 and up to 22 different resources. Only for 11 APTs, we collected a single resource, which typically is a white paper containing detailed information about the APT's activity over an extended period of time.}}}

We now answer~\ref{rq:quantitative_analysis} with a quantitative analysis of the attack vectors employed, the vulnerabilities exploited, and the software products targeted.

\subsection{Attack Vectors}
We analyzed the attack vectors exploited in the different campaigns with the presence and absence of software vulnerability.
Tab.~\ref{tab:table_attack_vectors} shows the different attack vectors and the
number of campaigns in which are observed. We underline that a
campaign can employ one or more attack vectors.\footnote{For example, it is not uncommon
to have campaigns that exploit both spearphishing and drive-by compromise.}
\begin{table}

	\caption{Attack vector campaigns and software vulns}
	\label{tab:table_attack_vectors}
	\centering
	\begin{adjustbox}{max width=\columnwidth}
		\begin{threeparttable}
			\begin{tabular}{l|r|r}
				\hline
				& \multicolumn{2}{c}{\bfseries \# of Campaigns}\\ \hline
				\bfseries Attack vector & \bfseries w/o vuln  & \bfseries w/ at least one
				vuln \\
				\hline\hline
				Spear phishing & 130*& 122*\\
				Drive-by Compromise & 15* & 34*\\
				Supply Chain Compromise & 5* & 0 \\
				Valid Accounts & 3* & 1 \\
				External Remote Services & 3 & 0 \\
				Exploit Public-Facing Appl. & 3* & 7 \\
				Replic. via Remov. Media & 0 & 1 \\
				Undetermined & 38* & 9* \\
				\hline
				Total & 197 (190 unique) & 174 (162 unique)\\
			\end{tabular}
			\begin{tablenotes}
				\item[*] Contains duplicates due to multiple attack vectors.
			\end{tablenotes}
		\end{threeparttable}
	\end{adjustbox}
\end{table}

We can observe that spear phishing is the main attack vector~\cite{urban13plenty}, present in 130 campaigns that do not
exploit any vulnerability and 122 campaigns that exploit at least one
vulnerability. Interestingly drive-by compromise is not only employed when a
vulnerability is present but also used to facilitate campaigns that employ
social engineering to trigger users to download malware.

We have 47 campaigns for which we do not know the attack vector. For 9 of them,
the report identified the vulnerability exploited but not the attack vector.\footnote{For example, some vulnerabilities (e.g.
CVE-2012-0158) can be exploited via spearphishing techniques and drive-by
compromise.} If this information is not present in the report, we avoided making
assumptions. For the remaining campaigns, the information about
the attack vector was vague or missing.\footnote{For example, the Sony hack
campaign in 2014~\cite{novettasonyhack}.}

\subsection{Popular Products and CVEs}
We observed 118 unique vulnerabilities exploited by the APTs in at least one
campaign between 2008 and 2020. Some CVEs are exploited in several campaigns by different APTs. 

Tab.~\ref{tab:table_product_campaigns} shows the ten most targeted client-side applications and the ten most targeted server/O.S. products based on the exploited CVEs. A campaign is counted over different products if the CVE employed is applicable to different software products. For example, CVE-2012-0158 affects Office, SQL server, Visual Fox Pro, and Commerce Server.\footnote{We do not have information about the exact software targeted. For example, they could all have exploited Office.} \textit{Office} is by far the major
target of campaigns followed by \textit{Windows O.S.} and \textit{Flash Player}. This is coherent with the attack vectors previously observed as they are commonly exploited via spearphishing with malicious attachments.
\begin{table}
	\caption{Top 10 client-side and Top 10 server-side/O.S. products exploited}
	\label{tab:table_product_campaigns}
	\centering
	\captionsetup[subfloat]{labelformat=empty}
	\subfloat[The products are obtained from the CVEs exploited in a campaign. If a CVE affects multiple products, all the software are considered. Products are distinguished in \emph{client-side}, \emph{server-side} application, and \emph{O.S.}]{
			\begin{tabular}{lp{0.31\columnwidth}p{0.12\columnwidth}r}
				\hline
				\bfseries Vendor & \bfseries Product & \bfseries Software & \bfseries \# Campaigns (\%)\\
				\hline      
				\textbf{Microsoft}&\textbf{Office}&\textbf{Client} & \textbf{68 (41.9\%)}\\
				Microsoft&Windows 2008 Server& O.S.& 49 (30.2\%) \\
				Microsoft&Windows 7& O.S. & 43 (26.5\%) \\
				Microsoft&Windows Vista&O.S. & 41 (25.3\%) \\
				Microsoft&Windows 2012 Server&O.S. & 39 (24.0\%) \\
				\textbf{Adobe} &\textbf{Flash Player (EOL)}& \textbf{Client} & \textbf{35 (21.6\%)} \\
				Microsoft&Windows 8.1&O.S. & 29 (17.9\%) \\
				Microsoft&Commerce Server&Server & 19 (11.7\%) \\
				Microsoft&SQL server&Server & 19 (11.7\%) \\
				Microsoft&Visual Basic&Client & 19 (11.7\%) \\
				Microsoft&Visual FoxPro&Client & 19 (11.7\%) \\
				Microsoft&BizTalk Server&Server & 18 (11.1\%) \\
				Microsoft&Windows 10&O.S. & 18 (11.1\%) \\
				Microsoft&Windows 8&O.S. & 14 (8.6\%) \\
				Microsoft&IE (EOL)&Client & 13 (8.0\%) \\
				\textbf{Adobe} &\textbf{Acrobat Reader}& \textbf{Client} & \textbf{11 (6.8\%)} \\
				Microsoft&.NET framework&Client & 5 (3.1\%) \\
				\textbf{Adobe} & \textbf{Air} & \textbf{Client} & \textbf{5 (3.1\%)} \\
				\textbf{Oracle} & \textbf{JRE} & \textbf{Client} & \textbf{4 (2.5\%)} \\
				Oracle &JDK&Client & 4 (2.5\%) \\
				\hline
			\end{tabular}
	}
\end{table} 

APTs tend to "share" vulnerabilities during their campaigns. Only 8 APTs (\texttt{Stealth Falcon}, \texttt{APT17}, \texttt{Equation},
\texttt{Dragonfly}, \texttt{Elderwood}, \texttt{FIN8}, \texttt{DarkHydrus}, and
\texttt{Rancor}) exploit CVEs that are not used by anyone else.\footnote{Only three APTs have exploited more than one vulnerability during all
	their campaigns.} We are aware of vulnerabilities (e.g
CVE-2017-0144) that are associated with \texttt{Equation} and used by other APTs, but
we did not find enough information about the date when the vulnerabilities were employed. Roughly 35\% of the APTs exploit CVEs observed in campaigns of other groups. 17 APTs share 4 or more vulnerabilities, while many APTs sharing a single vulnerability have only exploited that vulnerability during their campaigns (14 out of 20).

\subsection{Evolution in exploiting vulnerabilities}
\begin{figure}
	\captionsetup[subfloat]{labelformat=empty,width=\columnwidth}
	\centering
	\subfloat[The number of unique vulnerabilities employed in \emph{Unknown-Unknown (UU)} and \emph{Known-Unknown (KU)} attack scenarios grows
	significantly in recent years, compared to the first years of observation.
	On average around 5 distinct vulnerabilities per year are exploited
	by APTs.]
	{
		\includegraphics[width=0.8\columnwidth]{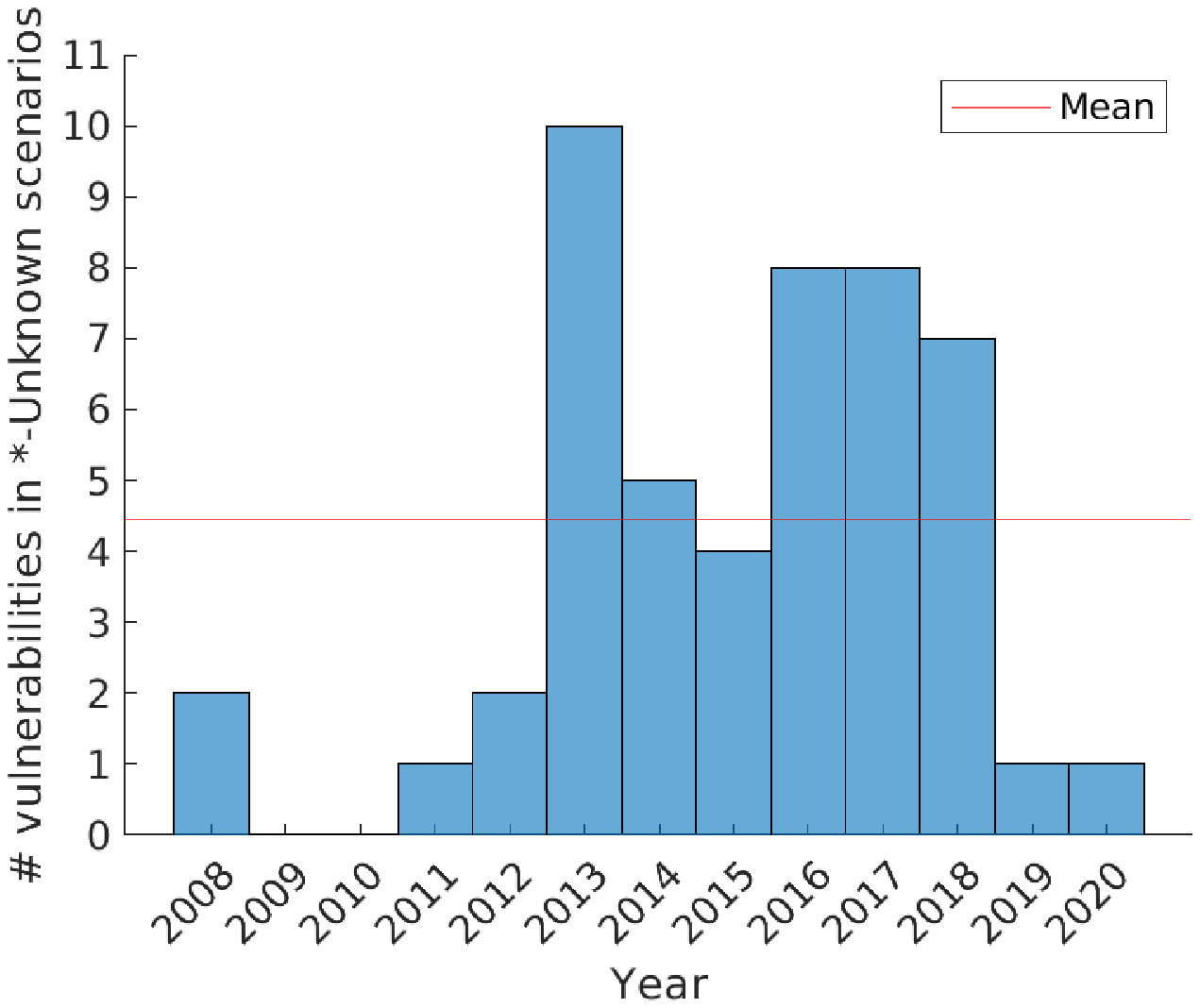}
	}
	\caption{Number of distinct vulnerabilities exploited over the years by different attack scenarios.}
	\label{fig:evolution-year-0days}
\end{figure}

Fig.~\ref{fig:evolution-year-0days} shows the evolution of the number of unique vulnerabilities exploited in the \emph{*-unknown} attack scenarios\footnote{Either already reserved \emph{(KU)} or not reserved \emph{(UU)}.} in our database. It represents a lower bound of the vulnerability exploited in the wild. Project Zero~\cite{projectzero_0-days} collects information about 0-days in the wild by including also unattributed attacks. The mean number of distinct vulnerabilities
exploited per year is roughly 5. We can observe how the numbers grew significantly in recent years. However, it can be
influenced by the limited number of reports for campaigns in the early period
(2008-2011), where it was less likely to report information about cyber-attacks. The drop for 2019/2020 is due to the natural delay of publicly reporting campaigns caused by the proximity of the period of data collection with the date of the campaigns themself. Thus, we expect the values to be higher if recomputed in the future.

Looking at the occurrence of a CVE in an APT campaign, the majority of the APTs prefer to exploit CVE already published, with few APTs as exceptions.\footnote{\texttt{Stealth Falcon, PLATINUM, APT17}}

\subsection{Software for Analysis of Update Strategies}\label{sec:sec:client-app}
As discussed in \S\ref{sec:methodology}, the collection of update releases from vendors' websites is a manual procedure. Here, for a first approximations, we focus on collecting updates for a subset of all software targeted by APTs.

Tab.~\ref{tab:table_product_campaigns} shows the most targeted products by vendor. We decided to cover the most exploited client-side product for each vendor because (1) from Tab.~\ref{tab:table_attack_vectors} most of the campaigns exploit attack vectors directed to client-side software and (2) it is not uncommon to have products from these vendors in an enterprise computer.
For Adobe, the \emph{Flash Player} product is end of life (EOL) thus we decided to include the other two software products \emph{Reader} and \emph{Air}. Even if \emph{Flash Player} is EOL in 2020 we still think it is interesting to see how different update strategies would affect the security of enterprises because it has been frequently exploited in the last years. Also, vendors' EOL of products, unfortunately, does not coincide with the disappearance from the field and end of exploitation as we are observing with Internet Explorer~\cite{secure_list_ie}.

For a first approximation, we limited the analysis to the \emph{Office} 2016 release only, as different releases (\emph{Office} 2013, \emph{Office} 365) can be seen as different products as they require buying a different license each. We considered the Knowledge Base (KB) updates from the Microsoft Update Catalog as the versions of the software. We assumed that KB updates for \emph{Office} are cumulative, i.e. the package contains all previously released fixes.

\noindent\fbox{\parbox{0.48\textwidth}{\noindent\emph{
In summary, we collected releases of updates for 5 different software products from 3 different vendors: \textit{Office}, \textit{Flash Player}, \textit{Acrobat Reader}, \textit{Air}, and \textit{JRE}.
We considered only releases for Microsoft Windows O.S. as it covers at least half of the enterprise computers~\cite{stats_windows}. 
With this set of software products, we cover 44\% of the campaigns (that exploit software vulnerabilities), 62\% of the APT groups, and 33\% of the CVEs.}}}

\section{Quantitative Analysis of Updates}\label{sec:analysis}
We now present an analysis of the speed of exploitation of individual vulnerabilities and the prevalence of \emph{*-Unknown} and \emph{Known-Known} attacks in APT campaigns. We then quantitatively evaluate the effectiveness and cost of the different update strategies against the APT campaigns.
\subsection{Survival Analysis}
\begin{figure}
	\captionsetup[subfloat]{labelformat=empty}
	\centering
	\subfloat[The survival is based on the \emph{first} time the CVE is exploited in a campaign. More than half of the vulnerabilities are exploited for the \emph{first time} within one month from the
	publication. However, there is high survivability of a small set of CVEs
	(roughly 10\%) that are exploited after more than 1 year from the publication. If we consider only \emph{Office}, \emph{Flash Player}, \emph{Reader}, \emph{Air}, and \emph{JRE} the behavior is similar.]
	{
		\includegraphics[width=\columnwidth,height=0.65\columnwidth]{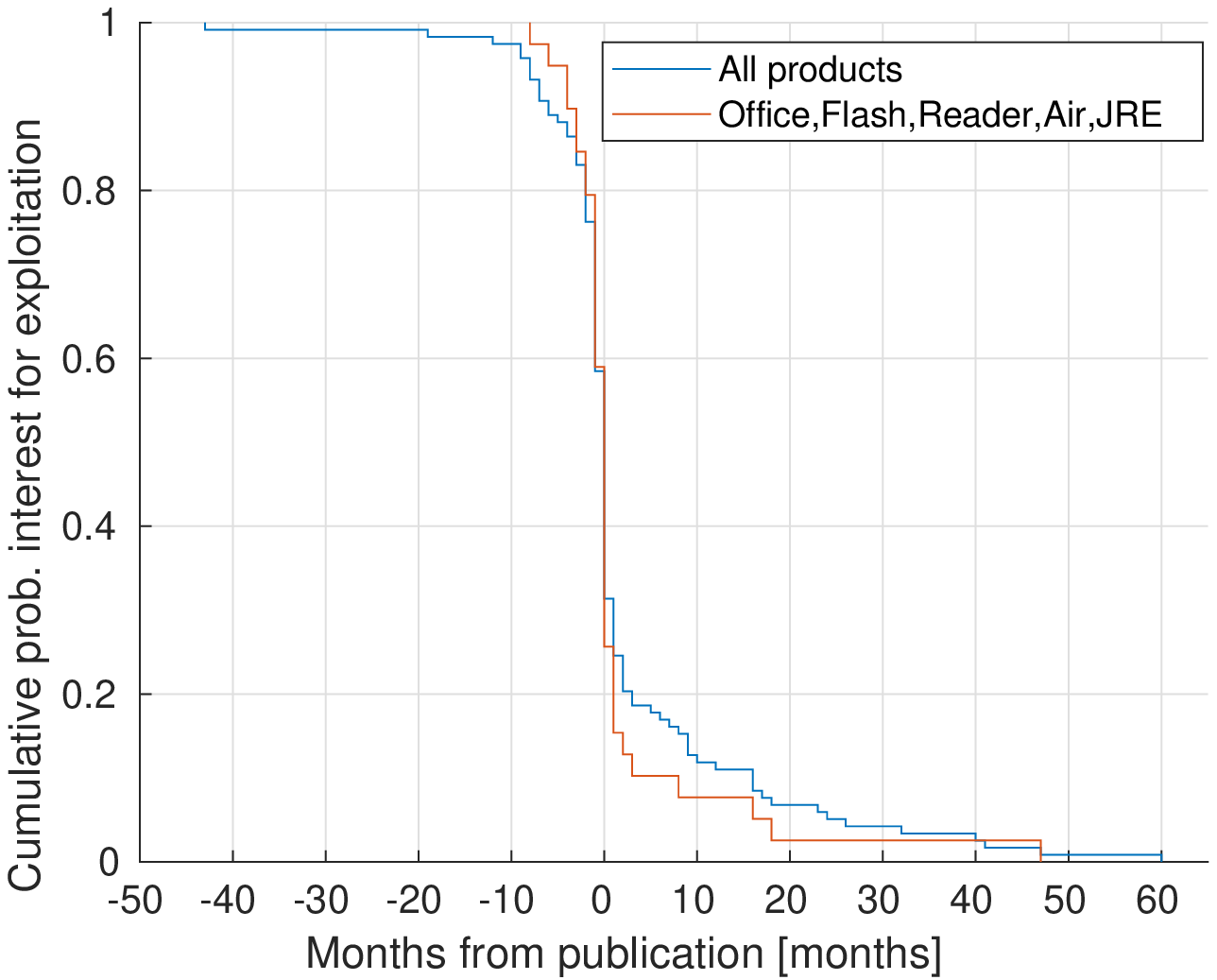}
	}
	\caption{Proportion of survival of CVE from publication (NVD) for all products and a subset (\emph{Office}, \emph{Flash Player}, \emph{Reader}, \emph{Air}, and \emph{JRE}). }
	\label{fig:distribution-delta-CVE-campaign}
\end{figure}

We performed preliminary survival analysis on the vulnerabilities to compute the interval in months that passed from the publication of the CVE
and the \emph{first} campaign that exploited the CVE (exploit age).
Fig.~\ref{fig:distribution-delta-CVE-campaign} shows the Kaplan-Meier plot for all the products in our database and for the set of products discussed in \S\ref{sec:dataset} (\emph{Office}, \emph{Flash Player}, \emph{Reader}, \emph{Air}, and \emph{JRE}). We
can see that roughly 40\% of the vulnerabilities are exploited for \emph{the first
	time} before the publication. This is coherent with what was observed by Chen et
al.~\cite{chen2019using}, where 49\% of the CVEs are exploited before the NVD
score is published. Furthermore, roughly 27\% of the vulnerabilities are
exploited \emph{the first time}\footnote{Among all the APTs.} within a month from the publication from NVD showing that APTs are fast to exploit new CVE~\cite{fireeye_vulns}. Another interesting fact is that a significant number of vulnerabilities are exploited a few months before the NVD publication. This phenomenon can be partially explained because the
observation of attacks in the wild brings software vendors to know about the vulnerability and thus the publication of a CVE.
It is important to underline that this value does not mean that $\approx$40\% of the campaigns are unpreventable because 1) \emph{*-Unknown} attacks can exploit several vulnerabilities\footnote{A famous example is Stuxnet.} and 2) many of these CVEs are exploited multiple times from different APTs after months from their first exploitation.

If we only consider the vulnerabilities exploited the first time in \emph{KK} attacks (as in~\cite{DBLP:conf/ccs/BilgeD12}), we observe that roughly 47\% of them are exploited within 30 days from their publications\footnote{Bilge et
al.~\cite{DBLP:conf/ccs/BilgeD12} observed a similar value of roughly 42\%}. In contrast with previous
results~\cite{DBLP:conf/raid/NayakMED14}, we observed a long tail for part of
the vulnerabilities, one out of 10 CVE is exploited after one year from its
publication, and 1 out of 20 after more than two years.

\subsection{Classification of APT campaigns}\label{sec:overview_campaigns}
Each APT campaign exploiting at least one vulnerability fits into one of these (possibly overlapping) groups:
\begin{itemize}
	\item Campaigns with at least one \emph{Known-Known (KK)} attack
	. In
	other words, the campaign exploited at least one vulnerability (either preventable or unpreventable) that was already present in the NVD database.
	\item Campaigns with at least one \emph{Known-Unknown (KU)} attack.
	In other words, the campaign exploited at least one vulnerability (either preventable or unpreventable) that was not present in the NVD database but an entry was already reserved by MITRE.\footnote{Thus, a small number of people known already some information about the vulnerability. E.g. vulnerability researchers.}
	\item Campaigns with at least one \emph{Unknown-Unknown (UU)} attack.
	In
	other words, the campaign exploited at least one vulnerability (either preventable or unpreventable) that was not even reserved
	by MITRE.
\end{itemize}

\begin{figure}
	\captionsetup[subfloat]{labelformat=empty}
	\centering
	\subfloat[The majority of campaigns exploited at least one vulnerability in a \emph{KK} attack (after publication by NVD and after reservation by MITRE). Only a few launched \emph{UU} attacks (both before reservation by MITRE and before publication by NVD).]{
		\includegraphics[width=0.9\columnwidth,height=0.6\columnwidth]{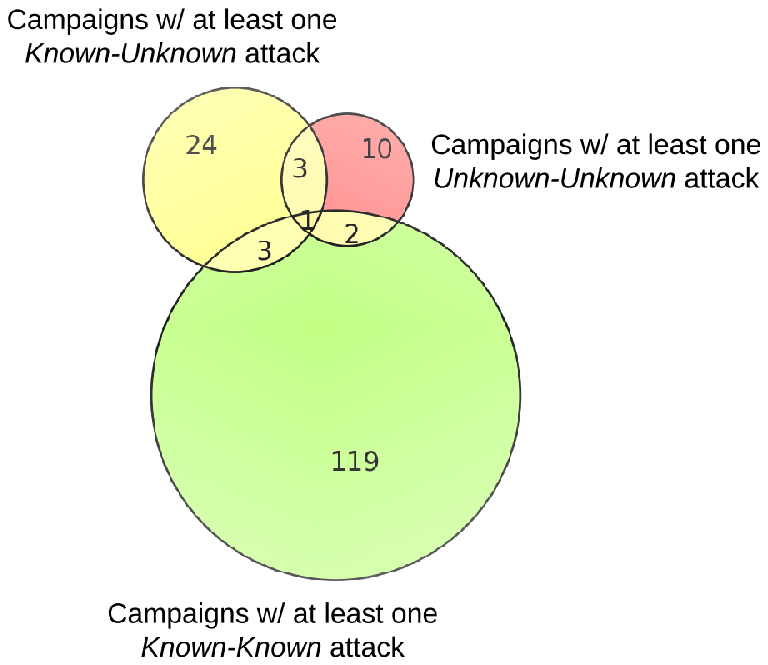}}
	\caption{Classification of APT Campaigns.}
	\label{fig:Venn-MITRE-NVD}
\end{figure}

Out of 352 campaigns, less than half of them employ at least one
vulnerability (Tab.~\ref{tab:table_attack_vectors}). 
Figure~\ref{fig:Venn-MITRE-NVD} shows the resulting Venn diagram for the 162
campaigns of interest. 119 out of 162 campaigns employed only vulnerabilities in \emph{Known-Known} attacks.
\noindent\fbox{\parbox{0.48\textwidth}{\noindent\emph{
APTs heavily exploit known CVE to compromise their target. The prioritization of updates is thus a key factor that can significantly reduce the impact of APTs campaigns.}
}}

\subsection{Evaluation of software updates strategies}

We now answer~\ref{rq:effectiveness} by applying our methodology (\S\ref{sec:methodology}) to compute the overall probability of being compromised (Eq.~\ref{eq:overall_prob_compromise}) in the interval of time [Jan 2008-Jan 2020] with the updates strategies and update interval presented in \S\ref{sec:terminology} for the software discussed in \S~\ref{sec:sec:client-app}. Tab.~\ref{tab:compromise_patchinterval} summarizes the results in terms of the number of updates required, the conditional probability and the odds ratio for the \emph{optimistic} (\emph{Update first}) and \emph{pessimistic} (\emph{APT first}) scenarios.
	
Updating the software as soon as a new release is available (\emph{Immediate} strategy) provides the optimal lower-bound probability of being compromised. Even in this case, roughly 1 out of 4 campaigns can compromise the target. Although an immediate update can be applied in some critical situations, if we consider a more realistic approach in which the software is updated with some delay in the month (\emph{Immediate} with \emph{APT first}), the odds of being compromised increases by a factor of 5.
	
The \emph{Planned} strategy provides a similar, although slightly better, probability of being compromised compared to a strategy that waits for the presence of public vulnerabilities (\emph{Reactive} strategy). However, waiting to update when a CVE is published presents 8x times fewer updates.
Thus, if an enterprise cannot keep up with the updates and need to wait before deploying them, can consider being simply reactive.
For the \emph{Planned} strategy the number of updates decreases with bigger intervals because the updates are shifted outside of the period of observation.
If a longer update interval is used, the probability of being compromised increases by a factor of ~9 and ~20 for 3 months and 7 months update intervals respectively. Interestingly, for the 7 months delay, we have that the \emph{Reactive} and \emph{Informed Reactive} perform slightly better than the \emph{Planned} strategy.
	
Comparing the \emph{Reactive} and \emph{Informed Reactive} strategies, there is a small advantage in knowing about not publicly known vulnerabilities only if the update interval is small. Once the enterprise waits 3 to 7 months, the vulnerability is now publicly known and actively exploited by the APTs.
	
\begin{figure*} 
	\centering
	\subfloat[\emph{Update-first} (optimist) scenario. There is a difference between the \emph{Immediate} and the other strategies. However, \emph{Planned}, \emph{Reactive}, and \emph{Informed Reactive} behave similarly thus updating to each new version with some delay or relying on reserved CVE does not worth it.]{
	\includegraphics[width=0.90\textwidth]{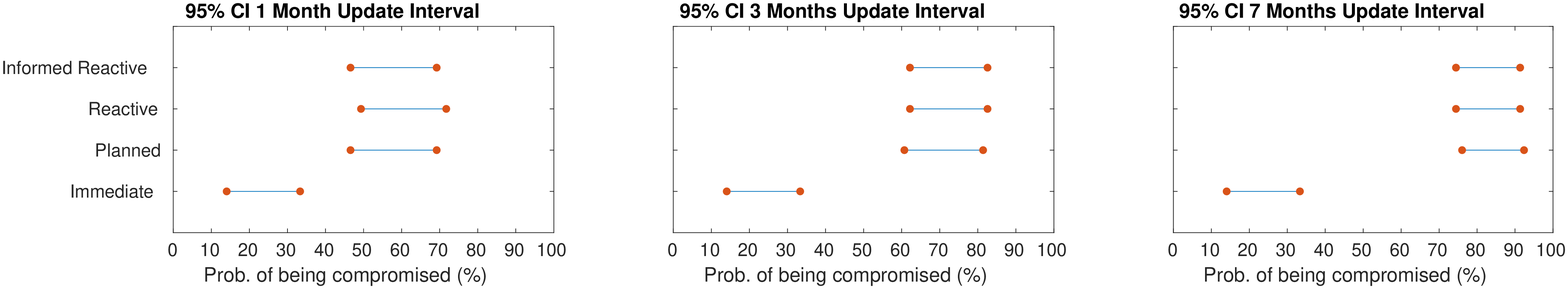}
}
	\\
		\subfloat[\emph{APT-first} (pessimist) scenario. The \emph{Immediate} and \emph{Planned} present a similar behavior for the 1 month update interval but differ with bigger intervals. In the pessimistic scenario the \emph{Planned}, \emph{Reactive}, and \emph{Informed Reactive} behave similarly thus updating to each new version with some delay or relying on reserved CVE does not worth it.]
	{%
		\includegraphics[width=0.90\textwidth]{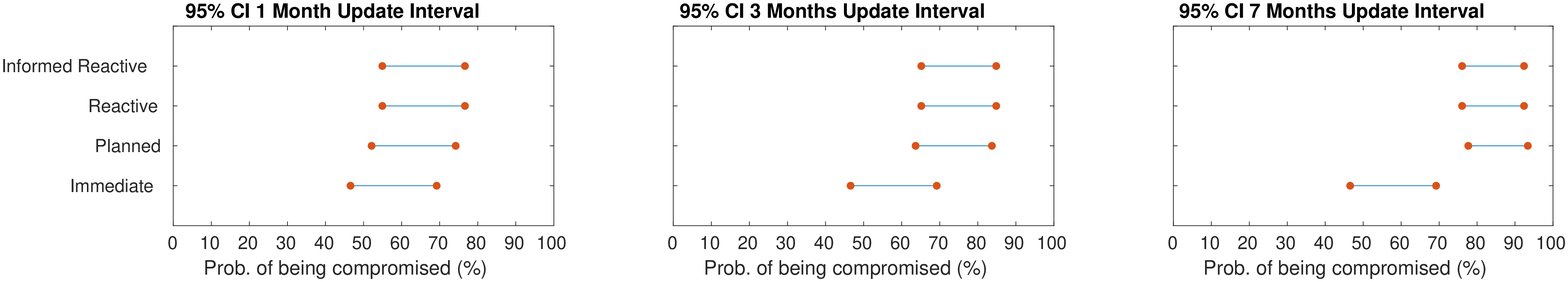}%
		\label{fig:ci_1month_neg}%
	}%
	
	\caption{Agresti-Coull Interval (CI) for the update strategies with different update intervals}
	\label{fig:ci_strategies}
\end{figure*}

We reported in Fig.~\ref{fig:ci_strategies} the Agresti-Coull Interval for each update strategy for the different update intervals. The \emph{Planned}, \emph{Reactive}, and \emph{Informed Reactive} strategies are almost identical as we see a significant overlap of the CI among these three strategies. The probability of being compromised lies within [52\%-74\%] for the \emph{Planned} and \emph{Informed Reactive} and [55\%-77\%] for the \emph{Reactive} in the pessimistic scenario. In the case of an optimistic \emph{Update-first} scenario, we observe that there is a clear difference between the \emph{Immediate} and the \emph{Planned} strategies, while this advantage is lost in the case of the pessimistic \emph{APT-first} scenario. To evaluate the similarity we computed, for each pair of strategies, the proportion of campaigns that either succeeded or failed against both strategies. We estimate the Agresti-Coull CI for the resulting proportions. The results show that the \emph{Planned} and \emph{Reactive} behave in the same way for at least 90\% up to 99\% of the campaigns for a 1 month update interval, for 88\% and up to 98\% for a 3 months update interval, and for 92\% up to 99\% for a 7 months interval. While, the \emph{Reactive} and \emph{Informed Reactive} behave in the same way for 90\% and up to 99\% of the campaigns for a 1 months update interval, and for 94\% up to 100\% for a 3 and 7 months interval.
	\noindent\fbox{\parbox{0.48\textwidth}{\noindent
	\emph{
Since hackers focus on new versions, a strategy that always updates to 
the new version but with a delay gives time to the APT to target and 
exploit a vulnerability. In contrast, a reactive approach that updates 
rarely might present to attackers an older version that does 
not include the new vulnerable code~\cite{DBLP:journals/tse/DashevskyiBM19}. 
In other words, either you update always 
\emph{and} immediately 
to the new versions or just updating lately has the same risk profile but cost you a lot more than updating rarely~\cite{DBLP:journals/ieeesp/MassacciJP21}.
	}}}

\begin{table}
	\captionsetup[subfloat]{labelformat=empty}
	\renewcommand{\arraystretch}{1.5}
	\caption{Optimistic (\emph{Update first}) and pessimistic (\emph{APT first}) overall \emph{conditional} probability of being compromised for different update strategies and update interval with the associated \# of updates for the period [01/2008-01/2020]}
	\label{tab:compromise_patchinterval}

	\centering
		\begin{tabular}{p{0.12\columnwidth}p{0.26\columnwidth}@{}rcc}
			\hline
			Update & Strategy  & \#Updates & Prob. & Odds \\
Interval & & & \multicolumn{2}{c}{\emph{(Update first --- APT first)}} \\
			\hline\hline
			/ & Immediate & 360 & 22.2-58.3\% &1x-4.9x\\ \hline
			\multirow{3}{*}{1 Month} & Planned & 357 & 58.3-63.9\% & 4.9x-6.2x \\
			& Reactive& 44 & 61.1-66.7\% & 5.5x-7.0x \\
			& Informed Reactive & 44 & 58.3-66.7\% & 4.9x-7.0x \\ \hline
			\multirow{3}{*}{3 Months} & Planned & 350 & 72.2-75.0\% & 9.1x-10.5x\\
			& Reactive & 44 & 73.6-76.4\% & 9.8x-11.3x \\
			& Informed Reactive & 44  & 73.6-76.4\% & 9.8x-11.3x\\ \hline
			\multirow{3}{*}{7 Months} & Planned & 337 & 86.1-87.5\% & 21.7x-24.5x \\
			& Reactive& 44 & 84.7-86.1\% & 19.4x-21.7x\\
			& Informed Reactive & 44  & 84.7-86.1\% & 19.4x-21.7x\\	
			\hline\hline
			\hline
		\end{tabular}
\end{table}

\section{Limitations}\label{sec:limitations}
The dataset obtained is based on publicly available reports.
While this is just a small part of existing campaigns, this paper is the 
first that tries to aggregate a manually validated dataset of APTs 
campaigns, CVE, and vulnerable products
and it is a first step in the direction of an open and extensive dataset 
on APT
campaigns. 

The process to obtain information about campaigns was semi-
automated but required manual effort to analyze and to extract the key information about campaign
dates, CVE, and attribution. We assume that this type of information 
reported by reputable security companies is not deliberately wrong, and our methodology strives to find multiple sources reporting the 
same campaign to control for possible errors.
Since keyword-based automated searches
(e.g.~\cite{DBLP:conf/esorics/LaurenzaL19}) present limitations in the number of
false associations that they generate, we decided that a manual approach would provide a more 
precise description of the APT ecosystem. Although the manual extraction of information from reports does not present difficulties, it can include erroneous matching of APT campaigns. To limit that, the manual analysis was performed by two researchers independently and inconsistencies were resolved by a third researcher.

We decided to ignore reports about campaigns where not enough information about the
start and attribution was available. Thus, it is possible that certain
vulnerabilities discussed in the reports are not included
in the dataset.

We applied a conservative approach in extracting information from different reports reporting mutually disjoint CVEs exploited on the same date. Thus, potentially assuming fewer campaigns with a higher number of CVEs each. The probability of being compromised must be seen as an upper bound of what APT can achieve. However, the odds ratio between update strategies remains the same.

We relied on the NVD data as the industry standard but it is known to contain errors in the
list of product names, CVE publication date~\cite{anwar2021cleaning} and vulnerable
versions~\cite{DBLP:conf/uss/DongGCXZ019,DBLP:journals/ese/NguyenDM16}. We leave for future work the application of these approaches to find inconsistencies. We relied on the data of
observation of the campaigns as reported in the reports we consulted. This information could be wrong and detect only a more recent campaign. We tried,
when possible, to find multiple resources about the campaign. The collection of release dates for the software discussed in \S\ref{sec:sec:client-app} is collected manually given that vendors’ repositories are not intended for past versions. Thus, the releases collected and employed in the evaluation might have errors and this could affect the \emph{Immediate} and \emph{Planned} strategies.

We used a month-based date granularity for the publication of the CVE, the
release of new versions, and the date of the campaigns because the exact day in a month in which the campaign started is not known. This decision has a
potential impact on the results. If a campaign for a CVE published on 29/01/2017 started on 01/02/2017 then in our case the exploit age is one
month, even if the CVE is exploited a few days after the publication. However,
the results we observed (e.g. exploit age of vulnerabilities) are coherent with
previous observations of attacks in the wild~\cite{DBLP:conf/ccs/BilgeD12}, thus
we think that the number of these cases is minimal and do not affect the results.

The same considerations apply to the results in
Tab.~\ref{tab:compromise_patchinterval}: if a
release is performed on 15/02/2019 and a
campaign exploiting the software is executed on 03/02/2019, the month
granularity would traduce both actions as performed on 02/2019. We thus considered two complementary scenarios: an \emph{optimistic} scenario (\emph{Update first}) and a \emph{pessimistic} scenario (\emph{APT first}). In the \emph{Update first} the example above will traduce in the defender be able to update before the execution of the campaign. While in the \emph{APT first} we assumed the opposite.

Finally, those companies that have an update interval that is less than a month will present a probability of being compromised that stays between the \emph{Immediate Update-first} and the \emph{Immediate APT-first} scenarios.

We assumed that a campaign will be carried on from the
date when the campaign started up to the end of the observation (i.e. 2020). This causes an inflation of the number of campaigns that are active at a given instant of time in Eq.~\ref{eq:prob_compromise}. However, we follow a conservative approach and assumed that if an APT has access to a vulnerability it will always be able to employ it given that one is under attack. We discuss extensions in the \S\ref{sec:conclusions}.

\section{Conclusions and Future Work}\label{sec:conclusions}

In this work, we proposed \emph{a methodology to quantitatively 
investigate the effectiveness and cost of software updates strategies 
against APT Campaign}. We applied the methodology to build  a 
database of APT campaigns and presented an analysis of the attack 
vectors, vulnerabilities, and
software exploited by 86 different APTs in more than 350 campaigns
over 12 years.
The database is publicly available on 
Zenodo~\cite{giorgio_di_tizio_2022_6514817}.

In contrast to expectation, we showed that preventive mechanisms like updates can influence the probability of being compromised by APT. However, software updates based on wrong measures of risk can be counterproductive.
Our analysis shows that a purely \emph{Reactive} update strategy (wait
until a vulnerability gets out) presents results very similar to a 
\emph{Planned} strategy (always update to the newest version), but with only 12\% of the updates. Furthermore, the \emph{Informed Reactive} strategy, where updates are applied based on
reserved information about not publicly known vulnerabilities (e.g. by
paying for information on 0-days), does not produce significant advantages compared to the \emph{Reactive} strategy and it is useless if the enterprise has several months of delay before applying the update.
\noindent\fbox{\parbox{0.48\textwidth}{\noindent
\emph{
In summary, for the broadly used products we analyzed, if you cannot keep updating always and immediately (e.g. because you must do regression testing before deploying an update), then being purely reactive on the publicly known vulnerable releases has the same risk profile than updating with a delay but costs significantly less.
}}}

Future work can extend the analysis to a more complete set of
software products and evaluate a subset of campaigns by targeted 
enterprises, attacker preferences, or network exposure based on IDS 
alerts~\cite{allodi2017security}. To achieve that, one would require to 
have company-specific information to move from a conditional probability to an absolute probability.

We also plan to extend the evaluation by considering campaigns as active only for a limited period. Further data about the lifetime of campaigns in the wild is required.

\ifCLASSOPTIONcompsoc
\section*{Acknowledgments}
\else
\section*{Acknowledgment}
\fi

We thank the student Veronica Chierzi for helping in the data collection of update releases.
This work was partly funded by the European Union under the H2020 Programme under grant n. 830929 (CyberSec4Europe) and n.952647 (AssureMOSS).

\bibliographystyle{IEEEtran-short}
\bibliography{short,references}

\end{document}

\title{Dataset and Replication Guide - Software Updates Strategies: a Quantitative Evaluation against Advanced Persistent Threats}

\author{Giorgio Di Tizio, Michele Armellini, 
        Fabio Massacci
        
\IEEEcompsocitemizethanks{\IEEEcompsocthanksitem G. Di Tizio (corresponding author) is with University of Trento, Italy.\protect\\
E-mail: giorgio.ditizio@unitn.it
\IEEEcompsocthanksitem M. Armellini is with University of Trento, Italy.
\IEEEcompsocthanksitem F. Massacci is with University of Trento, Italy and Vrije Universiteit Amsterdam, The Netherlands.}% <-this % stops an unwanted space
}

\IEEEtitleabstractindextext{%

\begin{abstract}
This is the replication guide for the paper \emph{Software Updates Strategies: a Quantitative Evaluation against Advanced Persistent Threats}
\end{abstract}

}

\maketitle

\IEEEdisplaynontitleabstractindextext

\IEEEpeerreviewmaketitle

\IEEEraisesectionheading{\section{Introduction}\label{sec:introduction}}

\IEEEPARstart{T}{he} supplementary materials in this paper provides additional information about the terminology used (\S\ref{appendix:terms}), the data collection procedure (\S\ref{appendix:procedure}), the structure of the database (\S\ref{appendix:neo4j_db}), the state-of-the-art on APTs (\S\ref{appendix:soa}), the dataset (\S\ref{appendix:APTlist} and \S\ref{appendix:multiple_vectors}), and the methodology applied to evaluate efficacy of update strategies (\S\ref{appendix:matrixexaple}).  

\section{Lists of Terms}\label{appendix:terms}
Tab.~\ref{tab:node_terminology}, ~\ref{tab:measure_terminology},
and~\ref{tab:definitions} contain the terminology that is used in the paper.
Some of these definitions are taken from the STIX
specification\footnote{\url{https://oasis-open.github.io/cti-documentation/resources\#stix-20-specification}.}.
\begin{table*}[h]
	\centering
	\caption[Nodes Terminology]{Nodes Terminology}
	\label{tab:node_terminology}
	\begin{tabular*}{\textwidth}{p{0.15\textwidth} p{0.8\textwidth}}
		\hline \hline
		\textit{APT} & A sophisticated group involved in malicious cyber
		activities.\\
		\textit{Vulnerability} &  A software flaw that can result in a
		security breach or a violation of the system's security policies.\\
		\textit{Campaign} & Time-bounded set of activity, carried out by an
		\textit{APT}, that uses particular techniques against a set of targets.\\
		\textit{Technique} & A method employed to achieve a specific goal
		like initial access, privilege escalation, etc.\\
		\textit{Product} &  A software product that is vulnerable to at
		least one vulnerability exploited by an \textit{APT}.\\
		\textit{Version} & A specific release of a software
		product that is vulnerable to at least one vulnerability exploited by an
		\textit{APT}.\\
		\textit{Country} &   Allegedly country of origin of an \textit{APT}.\\
		\hline
	\end{tabular*}
\end{table*}
\begin{table*}[h]
	\centering
	\caption[Measures Terminology]{Measures Terminology}
	\label{tab:measure_terminology}
	\begin{tabular*}{\textwidth}{p{0.2\textwidth} p{0.8\textwidth}}
		\hline \hline
		\textit{Reserved time $\reservedtime$} & Time when a CVE entry for a
		vulnerability is reserved by MITRE.\\
		\textit{Published time $\publishedtime$} & Time when the CVE for the
		vulnerability is published in NVD.\\
		\textit{Exploited time $\exploitedtime$} & Time when a campaign is
		being observed to start.\\
		\textit{Update Release time $\updatereleasetime$} & Time when the version of a software product is released.\\
		\textit{Observed time $t_{o}$} & Time when a
		product with a given version is observed to be vulnerable to
		a vulnerability. 
		\\
		\textit{Updated time $\mathit{t_{patched}}$}
		& Time when a software product is updated to a non vulnerable version.\\
		\textit{Exploit Age} & Interval of time between the publication of a CVE and the first
		observation of the vulnerability in a campaign.\\
		\hline
		\hline
	\end{tabular*}
\end{table*}

\begin{table}
	\caption{Common terms}
	\centering
	\begin{tabular}{p{0.30\columnwidth}p{0.60\columnwidth}}
		\hline
		\hline
		\textit{0-day vulnerability} & A vulnerability exploited w/o an update
		available.\\
		\textit{Unknown vulnerability} & A vulnerability exploited when it was not
		publicly known yet.\\
		\textit{Unknown attack vector} & The attack vector was either not
		specified in the report or was not known.\\
		\textit{CVE} & An unique identifier associated to a vulnerability by MITRE.\\
		\hline
		\hline
	\end{tabular}
	\label{tab:definitions}
\end{table}

\section{Procedure for data collection}\label{appendix:procedure}
We describe the manual procedure employed to collect data about APT campaigns and software updates from unstructured resources like technical reports and vendors' websites.
\subsection{Collection of APT campaigns}
We started from the list of APT groups obtained from MITRE~\cite{mitre_groups}. For each group:
\begin{enumerate}
	\item\label{it:first} we searched in the ThaiCERT Threat Group Cards v1.01~\cite{THAICERT-ENCICLOPEDIA} the resources describing campaigns associated with the group\footnote{At the time of writing a newer version (v2.0) is available at the ThaiCERT website.}.
	\item\label{it:second} we manually read each resource to identify the characteristics of the campaign in terms of date of execution, attack vector(s), and CVE(s) exploited. If a resource contains a date\footnote{With at least a month-base granularity.}, we created an entry that identify the campaign by its date of execution with additional information about the CVE exploited and the attack vector (technique): $<$\textit{APT\_name,CVE\_exploited,date\_start}$>$, $<$\textit{APT\_name,technique,date\_start}$>$. If multiple attack vectors/CVEs are employed in a campaign, we added one entry for each distinct attack vector/CVE.
	\item we searched for additional resources by web searches using as keywords: "\textit{APT\_name} CVE". We collected new resources until saturation was reached. For each new resource we reapplied~\ref{it:second}).
\end{enumerate} 

\subsection{Collection of Software updates}
We collected the information about the date and version of updates for 5 client-side applications: \emph{Office 2016}, \emph{Acrobat Reader}, \emph{Flash Player}, \emph{Air}, and \emph{JRE}.
We visited the official websites of the vendors (\emph{Microsoft}, \emph{Adobe}, and \emph{Oracle}) and searched for the web pages containing information about the releases and security updates for the products of interest. As vendors' repositories are highly unstructured and not intended for past version indexing we manually collected the date of release and version. In the case of old versions or EOL products, we relied on the Internet Archive\footnote{\url{https://archive.org/}} or external resources like JPCERT\footnote{\url{https://www.jpcert.or.jp/english/about/}} and Wikipedia.

\section{Neo4j Database}\label{appendix:neo4j_db}
We present the structure of the Neo4j database describing the nodes, their
properties, and their relationships.

We extended our dataset with the malware, tools, and techniques
associated with each APT using the \emph{pyattck} library~\cite{pyattack}. We then associated a list of aliases to each APT using the \textit{MISP ThreatActor galaxy}~\cite{misp_galaxy}.
\subsection{Node Labels}
\subsubsection*{APT}
This node label follows the lines of the STIX Threat Actor type. The information has been manually extracted from MITRE Att\&ck. The node contains
the following data:
\begin{itemize}
	\item labels: describes the type of APT. For example, possible values
	are activist, criminal, crime-syndicate,
	nation-state;
	\item name: is the name of the threat group as in MITRE Att\&ck;
	\item description: is a short description of the group;
	\item goals: are the goals of the APT. For example, financial gain,
	espionage, and sabotage;
\end{itemize}
\subsubsection*{Country}
This node label contains the following data:
\begin{itemize}
	\item name: is the name of the Country. It represents the Country in which an
	APT is allegedly to have origin/location according to the resources consulted.
	We did not make any assumption on the country origin of an APT because
	attribution is a well-known problem in the threat intelligence
	field.~\cite{AttributionGuerrero}
\end{itemize}
\subsubsection*{Alias}
This node label contains the following data:
\begin{itemize}
	\item name: is another name with which the APT is called. For example,
	APT 18\footnote{It is the name used by MITRE Att\&ck.}, is called
	\emph{Dynamite
		Panda} by Crowdstrike, \emph{Scandium} by Microsoft, \emph{Wekby} by Palo
	Alto,
	etc.;
\end{itemize}
\subsubsection*{Identity}
This node label represents a sector that is targeted by a certain APT.
The following parameters are used:
\begin{itemize}
	\item name: is the name of the sector. For example energy, defense,
	telecommunications, etc.;
\end{itemize}
\subsubsection*{Vulnerability}
This node label describes a vulnerability exploited by a certain APT.
The following properties are present:
\begin{itemize}
	\item name: is the CVE associated with the vulnerability. For example
	CVE-2016-4113;
	\item baseScore: is the CVSS Severity and Metrics Base Score of the CVE (v2 or
	v3);
	\item reservedDate: is the date  when the CVE has been \emph{reserved} by MITRE
	(in the
	format MM-YYYY);
	\item publishedDate: is the date when the CVE has been \emph{published} in NVD
	(in the format MM-YYYY). The date when NIST publishes a CVE could differ from
	the date when the CVE is reserved by MITRE~\cite{chen2019using};
\end{itemize}
\subsubsection*{Campaign}
This node label describes a campaign carried by a certain APT. It contains:
\begin{itemize}
	\item date\_start: is the date when the campaign was first observed;
\end{itemize}
We ignore the date of the end of the campaign due to the fact that this
information is not reliable, if not absent at all, in the reports analyzed.
\subsubsection*{Malware}
This node label describes the malware used by certain APTs. It contains:
\begin{itemize}
	\item name: is the name associated with the malware;
	\item platform: is the list of platforms that are vulnerable to the malicious
	software. For example Windows, Linux, macOS;
\end{itemize}
\subsubsection*{Tool}
This node label describes a legitimate tool available that is exploited by a
threat
actor during their campaigns. Each node contains:
\begin{itemize}
	\item name: is the name of the tool. For example, \emph{Winexe};
\end{itemize}
\subsubsection*{Technique}
This node label describes a technique used by an APT to compromise the
target.
It contains:
\begin{itemize}
	\item name: is the name of the technique as define in MITRE Att\&ck. For
	example, \emph{File and Directory Discovery};
	\item tactic: is the goal for which this technique is used. For example,
	\emph{initial access}, \emph{command-and-control}, etc.;
	\item platforms: is the platform affected. For example, Linux, Windows, macOS;
	\item permissions:  are the permissions required to implement this technique.
	For example, user, administrator, etc.;
\end{itemize}
\subsubsection*{Product}
This node label describes a software product that is vulnerable to at least a
CVE
exploited by an APT.
It contains the following property:
\begin{itemize}
	\item name: is the name of the product. For example, Internet Explorer;
\end{itemize}
\subsubsection*{Version}
This node label describes the version related to a specific software product for
which a
CVE has been published.
It contains the following properties:
\begin{itemize}
	\item name: is the version of the product;
	\item product: is the name of the product;
	\item update: is the name of the update, if any. For example \textit{sp1} for
	Windows XP;
	\item os: operating system(s) where the product can run;
\end{itemize}
\subsection{The Relationships}
The graph database presents different relationships that link nodes:
\begin{itemize}
	\item \textit{APT}$\xrightarrow{uses}$\textit{Malware}: defines which
	malware is used by a certain APT;
	\item \textit{APT}$\xrightarrow{uses}$\textit{Tool}: defines which
	tool is used by a certain APT;
	\item \textit{APT}$\xrightarrow{uses}$\textit{Technique}: defines
	which technique is used by a certain APT;
	\item \textit{APT}$\xrightarrow{origin}$\textit{Country}: defines the allegedly
	country of origin of an APT;
	\item \textit{APT}$\xleftarrow{attributed\_to}$\textit{Campaign}:
	describes a campaign allegedly to be attributed to a certain APT;
	\item Campaign$\xrightarrow{targets}$\textit{Vulnerability}: describes which
	CVE is exploited in a specific campaign;
	\item \textit{APT}$\xrightarrow{targets}$\textit{Identity}: defines
	which sector is targeted by a certain APT;
	\item \textit{APT}$\xleftarrow{alias}$\textit{Alias}: defines a
	relation between the names associated by different companies to a certain APT;
	\item Product$\xrightarrow{has}$\textit{Version}: defines which version are
	associated to a product;
	\item Version$\xrightarrow{vulnerable\_to}$\textit{Vulnerability}: defines
	which CVE affect a version of a product.
	\item Campaign$\xrightarrow{employs}$\textit{Technique}: defines
	which Technique is used to perform the \emph{initial access}. This is based on
	the MITRE Att\&ck Enterprise Initial Access section.
\end{itemize}

\section{State-of-the-art research questions}\label{appendix:soa}
Tab.~\ref{tab:PaperRQ} shows the research questions of the SoA and the related
results concerning APTs and TI in the different categories.
\begin{table*}\footnotesize
	\caption{Research Questions and Answers from the State of the Art about APTs over the years}
	\centering
	\begin{tabular}{p{0.05\columnwidth}p{0.05\columnwidth}p{0.2\columnwidth}p{0.6\columnwidth}p{1.0\columnwidth}}
		\hline
		\hline
		Paper & Year & Category & Research Questions & Answers\\ \hline
		\cite{giura2012context} & 2012 & Detection of attacks & 1)How can we model APT
		attacks? & An attack pyramid model with the goal on the top and the planes
		representing the environments. Events are placed in the pyramid.\\& & & 2)How can
		we detect APTs based on this model?&Correlated events describe an attack.
		Detection is based on signatures, profiling, and policy rules.\\ \hline
		\cite{zhao2014extended} & 2014 & Detection of attacks & 1)How can we model APT
		attacks?& An extended Petri net is developed to describe attack goals, the
		network, and the attacker's actions.\\ \hline
		\cite{DBLP:conf/cms/ChenDH14} & 2014 & Analysis of attackers characteristics &
		1)What are the characteristics of APT attacks?&APTs attack specific targets, are
		well-resourced, and stealthy.\\& & & 2)What are the phases of an APT attack?&
		Recon, initial intrusion, C\&C, lateral movement, and exfiltration.\\ \hline
		\cite{DBLP:conf/sose/BhattYG14} & 2014 & Detection of attacks & 1)How can we
		associate logs to phases of an APT attack?& A framework with an embedded
		intrusion kill chain receives logs from sensors and correlated events to
		determine phases of the attack\\ \hline
		\cite{DBLP:conf/icacci/ChandranPP15} & 2015 & Detection of attacks & 1)How can we
		detect APT malware on a system?& Using random forest to classify malware based
		on features like CPU and memory usage.\\ \hline
		\cite{DBLP:conf/infocom/HuLFCM15} & 2015 & Game Theory & 1)How can we characterize
		defense/attacker strategies with the presence of intruders?& A two-layer
		differential game between defender and attacker and multiple insiders. Nash
		equilibrium exists for both layers.\\ \hline
		\cite{DBLP:journals/compsec/FriedbergSSF15} & 2015 & Detection of attacks & 1)How
		can we detect phases of APT attacks from logs?& A rule-based model that
		correlates events from host security logs to extract anomalies from normal
		behaviors.\\ \hline
		\cite{DBLP:conf/cycon/MarchettiGPC16} & 2016 & Analysis of exploitation likelihood & 1)How to determine
		the hosts with the highest risk of being targeted by APT attacks?& A framework
		that analyzes both internal (network logs) and external (blacklist, social
		media) data to determine uncommon behaviors and exposure indicators.\\ \hline
		\cite{DBLP:journals/cn/MarchettiPCG16} & 2016 & Detection of attacks & 1)How can we
		identify suspicious hosts employed to exfiltrate data based on their network
		traffic?& A framework extracts features from network flow records to determine
		anomalous variations from normal behaviors.\\ \hline
		\cite{brogi2016terminaptor} & 2016 & Detection of attacks & 1)How can we detect
		phases of APT attacks from IDS data?&  A framework determine phases of attacks
		based on the information flow between events detected by IDS.\\& & & 2)How can we
		link different alerts to a single attack?& A tag is associated to an event from
		the IDS and its propagation underlines the complete chain of the attack.\\
		\hline
		\cite{DBLP:conf/ciss/UssathJ0M16}& 2016 & Analysis of attackers characteristics &
		1)What are the characteristics of APT attacks?& APTs employ different phases
		(initial comprom., lateral mov., C\&C, and exfilt.). They rely on spearphishing
		and known vulnerabilities.\\\hline
		\cite{DBLP:conf/acsac/PeiGSM0ZSZX16}& 2016 & Detection of attacks & 1)How can we
		detect phases of APT attacks from logs?& A weighted graph from system logs
		determines dense connections among events triggered by the attackers and sparse
		connections with benign events.\\ \hline
		\cite{DBLP:journals/fgcs/GhafirHPHHRA18}& 2018 & Detection of attacks & 1)How can
		we detect in real-time APT attacks based on IDS data?& A ML system composed of
		three modules in sequence: generation of alerts, clustering of alerts, and
		prediction of APT attacks based on the correlated alerts obtained in the
		previous module.\\ \hline
		\cite{DBLP:journals/compsec/LemayCMF18}& 2018 & APTs data sources & 1)What are the
		open-source resources about different APTs?& The majority of the resources come
		from industry reports. A list of resources for different APTs is provided.\\
		\hline
		\cite{riskAPT}& 2018 & Analysis of exploitation likelihood, Game Theory & 1)Does exist a Nash
		equilibrium in the allocation of response  resources against APTs to limit the
		loss?& Assuming attack and response strategy constant in time, a greedy
		algorithm is developed to determine the Nash equilibrium against lateral
		movements.\\ \hline
		\cite{DBLP:conf/cdc/SahabanduXC0LP18}& 2018 & Detection of attacks, Game Theory &
		1)How can we model Dynamic Information Flow Tracking to determine optimal
		strategies for the defender?& A game-theoretic model of DIFT and adversarial
		information flow is formulated to determine the equilibrium of both defender and
		attacker.\\ \hline
		\cite{DBLP:conf/ccs/ShuASSJHR18}& 2018 & Detection of Attacks & 1)How can we
		efficiently perform threat hunting from data logs and alerts?& Monitored events
		(process, files, network sockets, etc.) are described in a temporal
		computational graph. A graph database is generated and queried to extract
		attacks.\\ \hline
		\cite{DBLP:conf/sp/MilajerdiGESV19}& 2019 & Detection of attacks & 1)How can we
		detect in real-time APT campaigns from low-level event traces? & Information
		flows between file, process, etc. in a provenance graph are correlated and
		mapped to a corresponding TTPs description. \\ \hline
		\cite{DBLP:conf/uss/ShenS19}& 2019 & Detection of attacks & 1)How can we detect
		phases of attacks from IPS?& IPS alerts are converted in short sentences and
		their context is analyzed using word embedding technique.\\& & & 2)How can we
		detect changes in strategies by the attackers?& Cosine similarity is used to
		quantify embedding changes over time.\\ \hline
		\cite{DBLP:conf/uss/LiDPMVS19}& 2019 & Metrics for TI sources & 1)What are the
		metrics to evaluate TI data feeds?& Volume, differential contribution, exclusive
		contribution, latency, coverage, and accuracy.\\& & & 2)What are the major
		limitations of TI sources?& Coverage (different feeds have few overlaps) and
		false positive indicators.\\ \hline
		\cite{bouwman2020different} & 2020 & Metrics for TI sources & 1) What does paid TI
		consist of? & Indicators, reports, requests for information, portal for previous
		data.\\& &  & 2) How paid TI compares to open TI sources? & There is almost no
		overlap between paid TI and open TI and also between different paid TI.\\ & & & 3)
		How do customers use TI? & Network detection, situational awareness, SOC
		prioritization, and informing busines decisions.\\ \hline
		\cite{urban13plenty} & 2020 & Analysis of attackers characteristics & 1)What are the
		most used tactics employed by APTs? & 80\% of the attacks employ spearphising.\\
		& & & 2)Are companies leaking information exploitable for the campaigns? & 90\% of
		the companies leak data exploitable for phishing campaigns.\\ \hline
		\cite{DBLP:conf/sp/Hassan0M20}& 2020 & Detection of attacks & 1)How to detect APT attacks at run-time?  & Extended EDR to include provenance graph to reduce false alarms and log retention\\ \hline
		\cite{DBLP:conf/ndss/HanP0MS20} & 2020 & Detection of attacks & 1)How to detect APT attacks at run-time? & A provenance graph and clustering is used to detect anomalies.\\ \hline
		\cite{alsaheelatlas}& 2021 & Detection of attacks & 1)How to reconstruct APT steps from logs? & Causal graph extracted from logs are analyzed via NLP and ML techniques.\\
		\hline
		\hline
	\end{tabular}
	\label{tab:PaperRQ}
\end{table*}

\section{APT list}\label{appendix:APTlist}
Tab.~\ref{tab:apts_list} reports the list of APTs classified by their major goal. The APTs in bold target at least one of the products of interest (\emph{Office}, \emph{Flash Player}, \emph{Reader}, \emph{Air}, \emph{JRE}) and thus are considered in the evaluation.
\begin{table*}
	\caption{List of APTs in the dataset grouped by goal. APTs in bold are considered in the evaluation}
	\label{tab:apts_list}
	\begin{tabular}{l|p{0.9\textwidth}}
		\hline
		\bfseries Goal & \bfseries APTs\\\hline\hline
		espionage & \textbf{admin@338}, \textbf{APT12}, APT16, APT17, \textbf{APT18}, \textbf{APT19}, \textbf{APT28}, \textbf{APT29}, \textbf{APT3},
		APT30, \textbf{APT32}, APT33, \textbf{APT37}, APT39, \textbf{BlackOasis}, BRONZE BUTLER, Charming Kitten,
		\textbf{CopyKittens}, Dark Caracal, \textbf{Darkhotel}, DarkHydrus, Deep Panda, DragonOK,
		Equation, Gallmaker, Gamaredon Group, Gorgon Group, Group5, \textbf{Ke3chang}, \textbf{Lazarus
			Group}, Leafminer, \textbf{Leviathan}, \textbf{Lotus Blossom}, Magic Hound, \textbf{menuPass}, Molerats,
		\textbf{MuddyWater},  Naikon, \textbf{NEODYMIUM}, \textbf{OilRig}, \textbf{Patchwork}, PLATINUM, Poseidon Group,
		\textbf{PROMETHIUM}, Putter Panda, \textbf{Rancor}, \textbf{Scarlet Mimic}, Sowbug, \textbf{Stealth Falcon}, Stolen
		Pencil, Strider, Suckfly, \textbf{TA459}, \textbf{Threat Group-3390}, Thrip, \textbf{Tropic Trooper},
		\textbf{Turla}, \textbf{Winnti Group}, APT1, APT41, \textbf{Dust Storm}, Elderwood, Honeybee, Kimsuky,
		Night Dragon, PittyTiger, \textbf{Taidoor}, The White Company, WIRTE.\\\hline
		sabotage & APT33, \textbf{Dragonfly}, Dragonfly 2.0, Equation, \textbf{Lazarus Group}, Sandworm
		Team, TEMP.Veles. \\ \hline
		financial gain & APT38, \textbf{Carbanak}, \textbf{Cobalt Group}, Deep Panda, FIN10, FIN5,
		FIN6, FIN7, FIN8, GCMAN, Gorgon Group, \textbf{Lazarus Group}, \textbf{RTM}, SilverTerrier, FIN4,
		\textbf{Silence}.\\
		\hline
	\end{tabular}
\end{table*}

\section{Campaigns attack vectors and CVEs}\label{appendix:multiple_vectors}
Tab.~\ref{tab:table_map_vector_vuln} and Tab.~\ref{tab:table_map_vector_novuln} shows the matrix by pair of attack vectors for the campaigns with and without vulnerabilities. We did not observe more than two attack vectors exploited in a single campaign in our dataset.
The numbers on the diagonal line identify the number of campaigns that exploited
only a specific attack vector, while on the remaining entries we have the number
of campaigns that employed two different vectors. The matrices are symmetrical
with respect to the diagonal line. The number of unique campaigns can be
obtained from the sum of the elements on the diagonal line with either the
elements above or below the diagonal.

\begin{table}
	\caption{Attack vectors campaigns w/ software vulns}
	\label{tab:table_map_vector_vuln}
	\resizebox{\columnwidth}{!}{
		\begin{tabular}{l|r|r|r|r|r|r|r|r|}
			& \spheading{Spear phishing}& \spheading{Drive-by Compromise}&
			\spheading{Supply Chain Compromise}& \spheading{Valid Accounts}&
			\spheading{External Remote Services}& \spheading{Exploit Public-Facing
				Application} & \spheading{Replication Through Removable Media} &
			\spheading{Undetermined}\\
			\hline
			Spear phishing& 111 & 11 & & & & & & \\\hline
			Drive-by Compromise& 11 & 22 & & & & & & 1\\\hline
			Supply Chain Compromise& & & & & & & & \\\hline
			Valid Accounts& & & & 1& & & &\\\hline
			External Remote Services & & & & & & & &\\\hline
			Exploit Public-Facing Application & & & & & & 7 & & \\\hline
			Replication Through Removable Media & & & & & & &1 &\\\hline
			Undetermined & & 1 & & & & & &8\\
			\hline
		\end{tabular}
	}
\end{table}
\begin{table}

	\caption{Attack vectors campaigns w/o software vulns}
	\label{tab:table_map_vector_novuln}
	\resizebox{\columnwidth}{!}{\begin{tabular}{l|r|r|r|r|r|r|r|r|}
			& \spheading{Spear phishing}& \spheading{Drive-by Compromise}&
			\spheading{Supply Chain Compromise}& \spheading{Valid Accounts}&
			\spheading{External Remote Services}& \spheading{Exploit Public-Facing
				Application} & \spheading{Replication Through Removable Media} &
			\spheading{Undetermined}\\
			\hline
			Spear phishing&125 & 3 & 1 & & & 1 & & \\\hline
			Drive-by Compromise& 3&11 & & & & & & 1\\\hline
			Supply Chain Compromise&1 & &3 & 1 & & & & \\\hline
			Valid Accounts & & & & 3& & & & \\\hline
			External Remote Services & & & & &3 & & & \\\hline
			Exploit Public-Facing Application &1 & & & & &2 & & \\\hline
			Replication Through Removable Media & & & & & & &0 & \\\hline
			Undetermined & &1 & & & & & & 37\\
			\hline
	\end{tabular}}
\end{table}

Tab.~\ref{tab:table_CVE_preference} shows the 12 CVEs observed in the highest
number of campaigns. These results are confirmed by the CISA and FBI
report~\cite{cisa_top_cve}.
\begin{table}[!t]
	\captionsetup[subfloat]{labelformat=empty}
	\renewcommand{\arraystretch}{1.3}
	\caption{Top CVE with highest number of campaigns.}
	\label{tab:table_CVE_preference}
	\centering
	\begin{adjustbox}{max width=\columnwidth}
		\subfloat[Campaigns across different CVEs can overlap. For example,
		\emph{APT30} employed both CVE-2010-3333 and CVE-2012-0158 in a single
		campaign.]{
			\begin{tabular}{lrr}       
				\hline
				\bfseries CVE &  \bfseries Affected Product(s) & \bfseries \# Campaigns\\
				\hline      
				CVE-2012-0158 & Office, Visual Basic, SQL server,... & 18 \\
				CVE-2017-11882 &Office & 13 \\
				CVE-2017-0199 & Office, Windows Server & 11 \\
				CVE-2010-3333 & Office & 8 \\
				CVE-2015-1641 & Office, Sharepoint, Word & 7 \\
				CVE-2016-4117 & Flash player & 7 \\
				CVE-2015-2545 & Office & 6 \\
				CVE-2015-5119 & Flash player & 6 \\
				CVE-2014-1761 & Office, Sharepoint, Word  & 5 \\
				CVE-2017-8759 & .NET & 5\\
				CVE-2011-0611 & Flash player, Air, Reader,...  & 5\\
				CVE-2014-4114 & Windows Vista, 7, Server 2008,... & 5\\
				\hline
			\end{tabular}
		}
	\end{adjustbox}
\end{table} 

\section{Matrix of Strategies and Campaigns}\label{appendix:matrixexaple}
Tables~\ref{tab:immediatematrix} and~\ref{tab:exploitmatrix} present a fragment
of the matrix for the \emph{Immediate} strategy and the timeline of
campaigns of \textit{Acrobat Reader}.
\begin{table}[t]
	\caption{Fragment of \emph{Immediate} strategy matrix}
	\label{tab:immediatematrix}
	\begin{minipage}{\columnwidth} The boxed version might be zero or one depending on whether we consider the 
		update first or APT first criteria for updates happening within the month. 
	\end{minipage}
	\resizebox{\columnwidth}{!}{
		\begin{tabular}{lcccccccc}
			\hline
			& 05/12 & 06/12 & 07/12 & 08/12 & 09/12 & 10/12 & 11/12 &
			12/12 \\ \hline
			reader-9.3.3   & 	&  & 	&  & 	& 	& 	& 	\\
			reader-9.3.4   & 	& 	& 	& 	& 	& 	& 	& 	\\
			reader-9.4      & 	& 	& 	& 	& 	& 	& 	& 	\\
			reader-10       & 	& 	& 	& 	& 	& 	& 	& 	\\
			reader-10.0.1 & 	& 	& 	& 	& 	& 	& 	& 	\\
			reader-10.0.2 & 	& 	& 	& 	& 	& 	& 	& 	\\
			reader-10.0.3 & 	& 	& 	& 	& 	& 	& 	& 	\\
			reader-10.1    & 	& 	& 	& 	& 	& 	& 	& 	\\
			reader-10.1.1 & 	& 	& 	& 	& 	& 	& 	& 	\\
			reader-10.1.2 & 	& 	& 	& 	& 	& 	& 	& 	\\
			reader-10.1.3 & 1	& 1	& 1 & \fbox{0/1}		& 	& 	& 	& 	\\
			reader-10.1.4 & 	& 	& 	& 1	& 1	& \fbox{0/1} 	& 	&  \\
			reader-11      & 	& 	& 	&	& 	& 1	& 1	& 1 \\
			\hline
		\end{tabular}
	}
\end{table}
\begin{table}[t]
	\caption{Fragment of the campaign matrix timeline}
	\label{tab:exploitmatrix}
	\resizebox{\columnwidth}{!}{
		\begin{tabular}{lcccccccc}
			& 05/2012 & 06/2012 & 07/2012 & 08/2012 & 09/2012 & 10/2012 & 11/2012 &
			12/2012 \\ \hline
			reader-9.3.3 & 6	& 7	& 7	& 7	& 7	& 7	& 7	& 7 \\	
			reader-9.3.4 & 6	& 7	& 7	& 7	& 7	& 7	& 7	& 7	\\
			reader-9.4 & 5	& 6	& 6	& 6	& 6	& 6	& 6	& 6	\\
			reader-10 & 5	& 6	& 6	& 6	& 6	& 6	& 6	& 6	\\
			reader-10.0.1 & 5	& 6	& 6	& 6	& 6	& 6	& 6	& 6	\\
			reader-10.0.2& 4	& 5	& 5	& 5	& 5	& 5	& 5	& 5	\\
			reader-10.0.3& 	& 	& 	& 	& 	& 	& 	& 	\\
			reader-10.1& 	& 	& 	& 	& 	& 	& 	& 	\\
			reader-10.1.1& 	& 	& 	& 	& 	& 	& 	& 	\\
			reader-10.1.2& 	& 	& 	& 	& 	& 	& 	& 	\\
			reader-10.1.3& 	& 	& 	& 	& 	& 	& 	& 	\\
			reader-10.1.4& 	& 	& 	& 	& 	& 	& 	& 	\\
			reader-11 & 	& 	& 	& 	& 	& 	& 	&  \\
		\end{tabular}
	}
\end{table}
	The algorithm Alg.~\ref{alg:AlgProbability} shows the pseudo-code utilized to compute the conditional probability of being compromised in an instant of time $t_{i}$. For each of the update strategy matrices (e.g. Tab.~\ref{tab:immediatematrix}), we create a matrix for each campaign and perform an element-wise multiplication between them. The result is a matrix with the same dimension that has 1 in the entry in which the update strategy has a product version installed when the campaign exploited it. We then sum all the rows and normalize to 1 (because a campaign can target multiple products). The result is a vector in which each entry is 1 or 0 depending if at a specific $t_i$ the campaign had success or not.
	\begin{algorithm}\footnotesize
		\KwData{matrix\_updates,list\_campaigns}
		\KwResult{A vector containing the number of potentially successfull campaigns for all products for each instant of time $t_i$, the number of active campaigns for all products for each instant of time $t_i$}
		
		potentially\_successful\_campaigns = zeroes(2008:2020); /* $|C_{P_{z},\mathit{Ver_{x}}}|_{t_{i}}$ */\;
		active\_campaigns = zeros(2008:2020) /* $|C_{P_{z}}|_{t_i}$ */\;
		\For{\textit{campaign} in \textit{list\_campaigns}}{
			matrix\_campaign = \textit{create\_matrix(campaign)}\;
			active\_campaigns += \textit{normalize(\textit{sum\_rows(matrix\_campaigns)})}\; 
			result = matrix\_updates.*matrix\_campaign\;
			tmp\_successful\_campaigns = \textit{normalize(\textit{sum\_rows(result)})}\;
			potentially\_successful\_campaigns += tmp\_successful\_campaigns\;
		}
		\caption{Computation of Conditional Probability}
		\label{alg:AlgProbability}
	\end{algorithm}

\ifCLASSOPTIONcaptionsoff
\newpage
\fi

\section*{Acknowledgment}
We thank the student Veronica Chierzi for helping in the data collection of update releases. This work was partly funded by the European Union under the H2020 Programme under grant n. 830929 (CyberSec4Europe) and n.952647 (AssureMOSS).

\bibliographystyle{IEEEtran}
\bibliography{short,references}